\documentclass[reprint,
 superscriptaddress,
 amsmath,amssymb,
 aps,
prl,
]{revtex4-2}

\usepackage[utf8]{inputenc} 
\usepackage[T1]{fontenc}    
\usepackage{hyperref}       
\usepackage{url}            
\usepackage{booktabs}       
\usepackage{amsfonts}       
\usepackage{nicefrac}       
\usepackage{microtype}      
\usepackage{graphicx}
\usepackage{xcolor, etoolbox}
\usepackage{amsmath,amsfonts,amsthm} 
\usepackage{mathtools}
\usepackage{ragged2e}
\usepackage{braket}      
\usepackage[english]{babel} 
\usepackage{enumitem}
\usepackage{xspace}
\usepackage{dsfont}
\usepackage{bm}
\usepackage{bbm}
\usepackage{dcolumn}
\usepackage{mathrsfs}
\usepackage{todonotes}

\newcommand{\bTr}{\textbf{Tr}}
\newcommand{\Tr}{\text{Tr}}

\hypersetup{
    colorlinks,
    linkcolor=blue,
    citecolor=blue,
    urlcolor=blue,
    breaklinks=true 
}

\DeclarePairedDelimiter\abs{\lvert}{\rvert}

\begin{document}

\title{Information and majorization theory for fermionic phase-space distributions}

\author{Nicolas J. Cerf}
    \email{nicolas.cerf@ulb.be}
    \affiliation{Centre for Quantum Information and Communication, École polytechnique de Bruxelles, CP 165, Université libre de Bruxelles, 1050 Brussels, Belgium}
\author{Tobias Haas}
    \email{tobias.haas@ulb.be}
    \affiliation{Centre for Quantum Information and Communication, École polytechnique de Bruxelles, CP 165, Université libre de Bruxelles, 1050 Brussels, Belgium}

\begin{abstract}
We put forward several information-theoretic measures for analyzing the uncertainty of fermionic phase-space distributions using the theory of supernumbers. In contrast to the bosonic case, the anticommuting nature of Grassmann variables allows us to provide simple expressions for the Glauber $P$-, Wigner $W$-, and Husimi $Q$-distributions of the arbitrary state of a single fermionic mode. It appears that all physical states are Gaussian and, thus, can be described by positive or negative thermal distributions (over Grassmann variables). We then prove several fermionic uncertainty relations, including notably the fermionic analogs of the (yet unproven) phase-space majorization and Wigner entropy conjectures for a bosonic mode, as well as the Lieb-Solovej theorem and the Wehrl-Lieb inequality. Our central point is that, although fermionic phase-space distributions are Grassmann-valued and do not have a straightforward interpretation, the corresponding uncertainty measures are expressed as Berezin integrals, which take on real values and are physically relevant.
\end{abstract}

\maketitle

\textit{Introduction}---Pioneered by Heisenberg almost a century ago~\cite{Heisenberg1927}, the uncertainty principle for incompatible measurements in quantum theories has been stated and refined in various ways. Since the well-known second-moment uncertainty relations~\cite{Kennard1927,Robertson1929,Robertson1930,Schroedinger1930} do not fully capture the uncertainty encoded in a distribution, the uncertainty principle is nowadays often expressed in terms of entropies instead of variances, see~\cite{Everett1957,Beckner1975,Deutsch1983,Kraus1987,Maassen1988,Berta2010,Wehner2010,Coles2012} for discrete and~\cite{Bialynicki-Birula1975,Frank2012,Hertz2019,VanHerstraeten2021a} for continuous-variable systems (see also~\cite{Haas2021a,Haas2022b,Haas2024} for quantum fields). Entropic uncertainty relations are often stronger than their variance-based counterparts and, hence, are of great importance for many applications, \textit{e.g.}, to construct strong entanglement witnesses~\cite{Walborn2009,Saboia2011,Schneeloch2019,Haas2021b,Haas2022a} and to test the security of quantum cryptography protocols~\cite{Grosshans2004,Renes2009,Berta2010,Tomamichel2011,Furrer2012,Tomamichel2012}. 

Recently, even more general formulations of uncertainty relations in the framework of majorization theory have been put forward, see~\cite{Partovi2011,Puchala2013,Narasimhachar2016} for discrete and~\cite{Lieb2014b,Lieb2016,Lieb2021,VanHerstraeten2021b} for continuous-variable systems (see \textit{e.g.}~\cite{Nielsen1999,Haas2022d,Haas2022c} for applications in entanglement theory). Intuitively speaking, the theory of majorization imposes a preorder on the set of probability distributions, and the uncertainty relations pinpoint the distributions with \textit{least} disorder. In such formulations, entropic and second-moment relations are implied by a more fundamental majorization relation, highlighting the generality of this approach. 

As phase-space representations hold complete information about a given quantum state, it has been of particular interest to formulate such order relations for quasi-probability distributions covering phase space. So far, this has only been achieved for the Husimi $Q$-distribution--which is the measurement distribution obtained when projecting onto coherent states--for several degrees of freedom, including, \textit{e.g.}, a single bosonic mode and a single spin~\cite{Lieb2014b}. In contrast, majorization and entropic uncertainty relations for the Wigner $W$-distribution of a bosonic mode remain open conjectures~\cite{VanHerstraeten2021a,VanHerstraeten2021b}. 

While much effort has been devoted to constructing and analyzing information-theoretic measures in phase space for bosonic modes~\cite{Weedbrook2012,Serafini2017} and finite-dimensional systems~\cite{Nielsen2010,Wilde2013}, an information-theoretic description of fermionic modes, which are heavily constrained by Pauli's principle~\cite{Pauli1925}, is substantially less developed. Although fermionic phase-space representations have been analyzed in depth already two decades ago~\cite{Cahill1999}, the recently rising interest in fermionic systems has focused on Gaussian states~\cite{Hackl2021} and their entanglement properties~\cite{Banules2007,Zander2010,Friis2013,Friis2016,Debarba2020} (also in field theories, see, \textit{e.g.},~\cite{Casini2009,Calabrese2004,Calabrese2009,Haas2024}).

In this Letter, we explore various notions of uncertainty measures for a single fermionic mode. After constructing the sets of physical and coherent states, we show that all physical phase-space distributions are Gaussian (\textit{i.e.}, thermal states of positive or negative temperature). This radical simplification, brought about by Pauli's principle, allows us to prove several uncertainty relations for the Glauber $P$-, Wigner $W$-, and Husimi $Q$-distributions, as well as a complete set of majorization relations in fermionic phase space. Although the $P$-, $W$-, and $Q$-distributions are Grassmann-valued, the associated uncertainty relations involve real-valued entropies, hence are meaningful. For further remarks, see~\cite{SM}.

\textit{Notation}---We use natural units $\hbar = k_{\text{B}} = 1$ and write quantum operators (classical variables) with bold (regular) letters, \textit{e.g.}, $\boldsymbol{O} \, (O)$, respectively.

\textit{Single fermionic mode}---We consider a single fermionic mode described by Grassmann-valued mode operators $\boldsymbol{a}, \boldsymbol{a}^{\dagger}$ fulfilling the anti-commutation relations $\left\{\boldsymbol{a}, \boldsymbol{a}^{\dagger} \right\} = \mathds{1}, \left\{\boldsymbol{a}, \boldsymbol{a} \right\} = \left\{\boldsymbol{a}^{\dagger}, \boldsymbol{a}^{\dagger} \right\} = 0$. By Pauli's principle, the only two Fock states are the vacuum $\ket{0}$ and excited state $\ket{1}$, which form an orthonormal basis of the  Hilbert space $\mathcal{H}_2$ as $\braket{n | n'} = \delta_{n n'}$. The mode operators act as ladder operators $\boldsymbol{a}^{\dagger} \ket{n} = \sqrt{n+1} \ket{n+1}, \boldsymbol{a} \ket{n} = \sqrt{n} \ket{n-1}$ for $n \in \{0,1 \}$. Denoting by $\braket{n} = \bTr \left\{ \boldsymbol{\rho} \, \boldsymbol{a}^{\dagger} \boldsymbol{a} \right\} \in [0,1]$ the total particle number allows us to write the most general single-mode fermionic density operator as
\begin{equation}
    \boldsymbol{\rho} = (1 - \braket{n}) \, \boldsymbol{a} \boldsymbol{a}^{\dagger} + \lambda \, \boldsymbol{a} + \lambda^* \, \boldsymbol{a}^{\dagger} + \braket{n} \boldsymbol{a}^{\dagger} \boldsymbol{a}.
    \label{eq:DensityOperator}
\end{equation}
with $\lambda \in \mathbb{C}$ and $\abs{\lambda} \le \sqrt{\braket{n} (1-\braket{n})}$ to ensure $\boldsymbol{\rho} \ge 0$.

\textit{Physical states and Gaussianity}---It has been argued that any \textit{physical} fermionic density operator is constrained by an additional superselection rule, which can be motivated by the spin-statistics theorem in relativistic quantum field theories~\cite{Hegerfeldt1968,Friis2013,Friis2016}: in Lorentz-invariant theories, fermions carry half-integer spin, and hence spatial rotations by $2\pi$ change a state with an odd (even) number of fermions by a factor of $-1$ ($+1$). Since the state needs to be invariant (up to a global phase) under such a rotation, physical states \textit{cannot} contain superpositions of odd and even particle numbers, which results in the requirement $\lambda=0$. Interestingly, this immediately implies that all physical states are thermal. Indeed, using the identity $y^{\boldsymbol{a}^{\dagger} \boldsymbol{a}} = \mathds{1} + (y-1)\,\boldsymbol{a}^{\dagger} \boldsymbol{a}$ for $y\ge 0$, any physical state ($\lambda=0$) can be written as~\cite{SM}\textbf{(a)}
\begin{equation}
    \begin{split}
        \boldsymbol{\rho} &= (1 - \braket{n}) \,\boldsymbol{a} \boldsymbol{a}^{\dagger} + \braket{n} \boldsymbol{a}^{\dagger} \boldsymbol{a} = \frac{1}{1 + e^{\nu}} \, e^{\nu \boldsymbol{a}^{\dagger} \boldsymbol{a}},
    \end{split}
    \label{eq:PhysicalDensityOperator}
\end{equation}
with $e^{\nu} = \braket{n}/(1-\braket{n})$. Thus, physical states are nothing but (Gaussian) thermal states with $\nu = - \epsilon/T$, where $T$ denotes the temperature and $\epsilon$ is the excitation energy. 

\begin{figure}[t!]
    \centering
    \includegraphics[width=0.85\columnwidth]{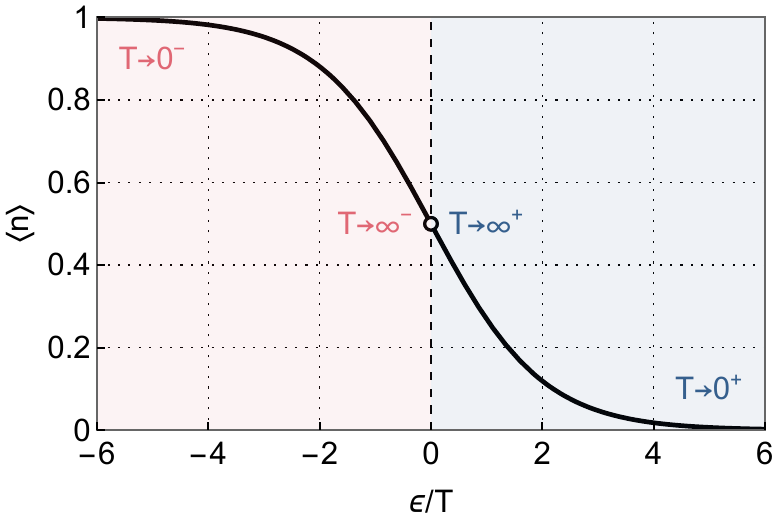}
    \caption{Physical single-mode fermionic states are thermal Gaussian states characterized by the Fermi-Dirac distribution $\braket{n}=1/(1+ e^{\epsilon/T})$. The positive (blue) and negative (red) temperatures are associated with $\braket{n} < 1/2$ and $>1/2$, with the extremal points $\ket{0}$ and $\ket{1}$ being reached in the limit $T \to 0^+$ and $T \to 0^-$, respectively.}
    \label{fig:FermiDiracDistribution}
\end{figure}

It is instructive to describe physical states in terms of the occupation number $\braket{n} = e^{\nu}/(e^{\nu} + 1) = 1/(1 + e^{\epsilon/T})$, which corresponds to the Fermi-Dirac distribution (see Fig.~\hyperref[fig:FermiDiracDistribution]{1}). Note first that the purity of~\eqref{eq:PhysicalDensityOperator} is $\bTr \{ \boldsymbol{\rho}^2 \} = 1 - 2 \braket{n} \left( 1 - \braket{n}\right)$, implying that the only two pure states are the vacuum ($\braket{n}=0$) and excited state $(\braket{n}=1)$. The family of physical states can be divided into positive- and negative-temperature thermal states. The two branches are connected by the maximally-mixed state with $\braket{n}=1/2$, which requires an infinite temperature of arbitrary sign $T \to \infty^{\pm}$. The vacuum and excited states correspond to the zero-temperature limits $T \to 0^+$ and $T \to 0^-$, respectively. They correspond to the extremal points of the set of physical states, and we may expect their uncertainty in phase space to play a special role, too.

\textit{Fermionic coherent states}---Following~\cite{Cahill1999}, we introduce the fermionic displacement operator $\boldsymbol{D}(\alpha) = e^{\boldsymbol{a}^{\dagger} \alpha - \alpha^* \boldsymbol{a}}$, with $\alpha,\alpha^*$ being Grassmann-valued variables such that $\left\{ \alpha, \alpha^* \right\} = \left\{ \alpha, \alpha \right\} = \left\{ \alpha^*, \alpha^* \right\} = 0$, which also anticommute with the Grassmann-valued mode operators $\boldsymbol{a}, \boldsymbol{a}^{\dagger}$, namely $\left\{ \alpha, \boldsymbol{a} \right\} = \left\{ \alpha, \boldsymbol{a}^{\dagger} \right\} = \left\{ \alpha^*, \boldsymbol{a} \right\} = \left\{ \alpha^*, \boldsymbol{a}^{\dagger} \right\} = 0$. It is easy to check that $\boldsymbol{D}(\alpha)$ is a unitary operator and $\boldsymbol{D}^\dagger(\alpha)=\boldsymbol{D}(-\alpha)$. Then, fermionic coherent states are defined as displaced vacuum states
\begin{equation}
    \ket{\alpha} = \boldsymbol{D} (\alpha) \ket{0},
    \label{eq:CoherentStates}
\end{equation}
with Grassmann-valued displacements $\alpha$. Although they are unphysical, the fermionic coherent states provide the right tool to express quasi-probability distributions in phase space. As in the bosonic case, the coherent state basis is overcomplete with the identity being represented as a Berezin integral~\cite{SM}\textbf{(b)}, $\mathds{1} = \int \mathcal{D} \alpha \boldsymbol{\ket{\alpha} \bra{\alpha}}$, where we have chosen the standard sign convention that the innermost integral is being performed first, \textit{i.e.}, $\int \mathcal{D} \alpha \, \alpha \alpha^* \equiv \int \mathrm{d} \alpha^* \mathrm{d} \alpha \, \alpha \alpha^* \equiv +1$. This also leads us to express the trace of an operator $\boldsymbol{O}$ in the coherent-state basis as $\bTr \{ \boldsymbol{O} \} = \int \mathcal{D} \alpha \braket{\alpha | \boldsymbol{O} | - \alpha}$. Note the discrepancy (minus sign) with the analog formula for bosonic coherent states. See also~\cite{SM}\textbf{(c)} for more details.

\textit{Phase-space distributions and supernumbers}---Apart from further minus signs ensuring correct normalizations, the phase-space distributions are defined analogously to the bosonic case. Hence, the Glauber $P$-distribution represents the diagonal elements of the state in the coherent-state basis, namely
\begin{equation}
    \boldsymbol{\rho} = \int \mathcal{D} \alpha \, P (\alpha) \, \boldsymbol{\ket{\alpha} \bra{-\alpha}},
    \label{eq:GlauberPDistribution}
\end{equation}
and the Wigner $W$-distribution is the Fourier transform of the characteristic function $\chi(\alpha)= \bTr \{ \boldsymbol{\rho} \, \boldsymbol{D}(\alpha) \}$, namely
\begin{equation}
    W (\alpha) = \int \mathcal{D} \beta \, e^{\alpha \beta^* - \beta \alpha^*} \, \chi(\beta),
\label{eq:WignerWDistribution}
\end{equation}
while the Husimi $Q$-distribution is the outcome distribution obtained when measuring $\boldsymbol{\rho}$ in the coherent-state basis, \textit{i.e.}
\begin{equation}
    Q (\alpha) = \bTr \{ \boldsymbol{\rho} \boldsymbol{\ket{\alpha} \bra{\alpha}} \} = \braket{\alpha | \boldsymbol{\rho} | -\alpha}.
    \label{eq:HusimiQDistribution}
\end{equation}
All three are Grassmann-even for physical states and can be computed analytically in full generality, which leads to the simple Gaussian expressions
\begin{equation}
    \begin{split}
        P (\alpha) &= - \braket{n} e^{-\frac{\alpha \alpha^*}{\braket{n}}} = - \braket{n} + \alpha \alpha^*, \\
        W(\alpha) &= \left( 1/2 - \braket{n} \right) e^{\frac{\alpha \alpha^*}{1/2 - \braket{n}}} =  1/2 - \braket{n} + \alpha \alpha^*, \\
        Q(\alpha) &= \left( 1 - \braket{n} \right) e^{\frac{\alpha \alpha^*}{1 - \braket{n}}} = 1 - \braket{n} + \alpha \alpha^*.
    \end{split}
    \label{eq:PhaseSpaceDistributions}
\end{equation}
Since the prefactor of the term $\alpha \alpha^*$ is one in all cases, the phase-space distributions are normalized to unity with respect to the Berezin integral measure $\int \mathcal{D} \alpha \, P(\alpha) = \int \mathcal{D} \alpha \, W(\alpha) = \int \mathcal{D} \alpha \, Q(\alpha) = \bTr \{ \boldsymbol{\rho} \} = 1$. Except for normalization, neither $P(\alpha)$ nor $W(\alpha)$ nor $Q(\alpha)$ has a straightforward physical interpretation (unlike their bosonic counterparts) since these are distributions over Grassmann variables, see also~\cite{SM}\textbf{(d)}.

To provide a deeper understanding of expressions~\eqref{eq:PhaseSpaceDistributions}, we make an excursion into the theory of supernumbers (see~\cite{DeWitt1992}). Every Grassmann number $z$, \textit{i.e.}, every element of the Grassmann algebra, is a supernumber and can be decomposed linearly as $z = z_{\text{B}} + z_{\text{S}}$. Therein, the so-called \textit{body} $z_{\text{B}} \in \mathbb{C}$ is the ordinary scalar part, while the so-called \textit{soul} $z_{\text{S}}= c_1 \alpha + c_2 \alpha^* + c_3 \alpha \alpha^*$ contains all Grassmann-valued contributions with complex-valued coefficients $c_i \in \mathbb{C}$. A supernumber $z$ is real if and only if $z^* = z$ and positive (negative) if and only if its body $z_{\text{B}}$ is positive (negative). The latter implies an ordering relation for supernumbers, namely, that $z_1 \le z_2$ if and only if $z_{1,\text{B}} \le z_{2, \text{B}}$ and vice versa.

Thus, the three phase-space distributions in~\eqref{eq:PhaseSpaceDistributions} are real and have equal (Grassmann-even) souls $P_{\text{S}} = W_{\text{S}}  = Q_{\text{S}}  = \alpha \alpha^*$, but different bodies $P_{\text{B}} = -\braket{n}$, $W_{\text{B}} = 1/2 - \braket{n}$, $Q_{\text{B}} = 1 - \braket{n}$. Hence, the Husimi $Q$-distribution is always larger than the Wigner $W$-distribution, which itself is always larger than the Glauber $P$-distribution, \textit{i.e.}, $Q(\alpha) > W (\alpha) > P(\alpha)$ for all $\alpha$ and $\braket{n}$, a relation which does not exist in the bosonic case. Furthermore, the Glauber $P$-distribution (Husimi $Q$-distribution) is always negative (positive) since $P_{\text{B}} \le 0$ ($Q_{\text{B}} \ge 0$) for all $\braket{n}$, while the Wigner $W$-distribution is entirely positive (or entirely negative) for $0 \le \braket{n} < 1/2$ (or for $1/2 < \braket{n} \le 1$), for which we write $W^+$ and $W^-$, respectively.

\textit{Majorization relations}---We generalize the definition of a majorization relation for a bosonic mode (see~\cite{Marshall2011,VanHerstraeten2021b}) straightforwardly: A phase-space distribution $z_1(\alpha)$ is said to be majorized by another distribution $z_2(\alpha)$, written as $z_1 \prec z_2$, if
\begin{equation}
    \int \mathcal{D} \alpha \, f (z_1) \ge \int \mathcal{D} \alpha \, f (z_2),
    \label{eq:MajorizationRelation}
\end{equation}
for all concave functions $f: \mathcal{I} (z_\text{B}) \to \mathbb{R}$ with $f(0) = 0$, where $\mathcal{I} (z_\text{B}) \subset \mathbb{R}$ denotes the image of $z_{\text{B}}=z_{\text{B}}(\braket{n})$. For the three phase-space distributions of interest, we shall prove the fundamental majorization relations (here, the index of $P$, $W$, or $Q$ refers to the value of $\braket{n}$):
\begin{enumerate}
    \item Any Glauber $P$-distribution is majorized by (majorizes) the vacuum (excited) state:
    \begin{equation} 
        P_1 \prec P \prec P_0.
        \label{eq:MajorizationGlauberP}
    \end{equation}
    \item Any Wigner $W$-distribution is majorized by (majorizes) the vacuum (excited) state, with the maximally-mixed state being majorized by (majorizing) all Wigner-positive (Wigner-negative) distributions:
    \begin{equation} 
        W_1 \prec W^- \prec W_{1/2} \prec W^+ \prec W_0.
        \label{eq:MajorizationWignerW}
    \end{equation}
    \item Any Husimi $Q$-distribution is majorized by (majorizes) the vacuum (excited) state:
    \begin{equation} 
        Q_1 \prec Q \prec Q_0.
        \label{eq:MajorizationHusimiQ}
    \end{equation}
\end{enumerate}
We stress that the rightmost majorization relations in Eqs.~\eqref{eq:MajorizationWignerW} and~\eqref{eq:MajorizationHusimiQ} resemble the bosonic phase-space majorization conjecture~\cite{VanHerstraeten2021b} and the Lieb-Solovej theorem~\cite{Lieb2014b,Lieb2016,Lieb2021}, respectively, while no bosonic analog of Eq.~\eqref{eq:MajorizationGlauberP} exists.

\begin{figure*}[t!]
    \centering
    \includegraphics[width=0.995\textwidth]{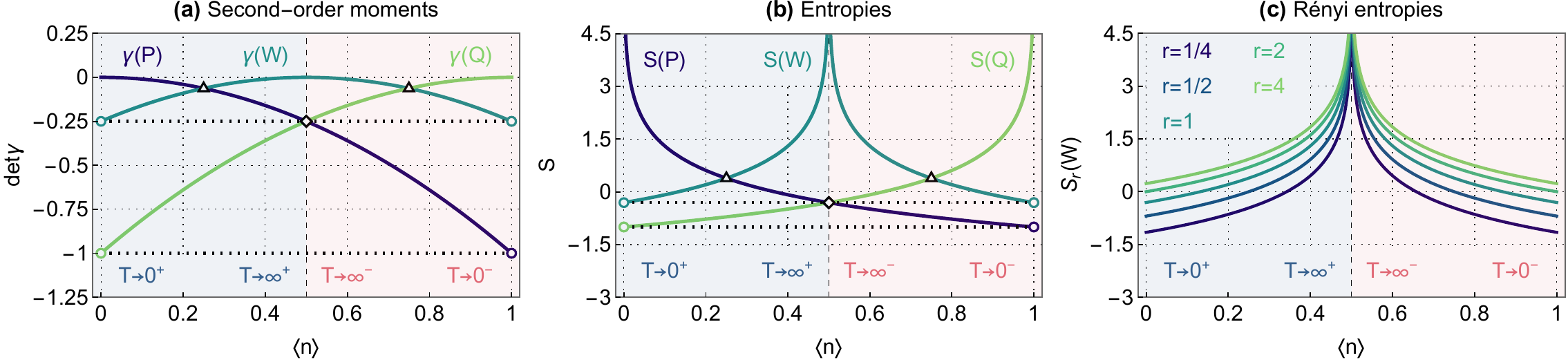}
    \caption{Measures of disorder in fermionic phase space and their lower bounds (black dashed) expressing the uncertainty principle as functions of the particle number $\braket{n}$. \textbf{(a)} shows the three second-moment measures, $\det \gamma (P)$ (purple), $\det \gamma (W)$ (petrol), and  $\det \gamma (Q)$ (green), while \textbf{(b)} displays their entropic analogs. The curves for the $W$- and $P(Q)$-distributions intersect at $\braket{n}=1/4$ ($3/4$) (triangles), while the $P$- and $Q$-distributions meet at $\braket{n}=1/2$ (diamonds), such that the $P$-, $W$- and $Q$-curves are the largest when $\braket{n} \le 1/4$, $1/4 \le \braket{n} \le 3/4$ and $\braket{n} \ge 3/4$, respectively. The extension to Rényi entropies for various entropic orders $r$ is presented in \textbf{(c)} for the Wigner $W$-distribution, showing that $S_r(z)$ is monotonically increasing in $r$.}
    \label{fig:UncertaintyRelations}
\end{figure*}

Their proofs rely on the central observation that concave averages in fermionic phase space are real numbers, which can be computed explicitly for all phase-space distributions $z=P,W^+,W^-,Q$ of interest. To show this, we use the notion of an analytic functional over a supernumber $z = z_{\text{B}} + z_{\text{S}}$ via its Taylor expansion~\cite{DeWitt1992}
\begin{equation}
    f (z) = \sum_{j=0}^{\infty} \frac{1}{j!} \,f^{(j)} (z_{\text{B}}) \, z_{\text{S}}^j,
    \label{eq:FunctionalOverSupernumbers}
\end{equation}
where $f^{(j)} (z_{\text{B}})$ denotes the $j$th derivative of $f$ evaluated at the body $z_{\text{B}}$. Using that the soul of all physical phase-space distributions~\eqref{eq:PhaseSpaceDistributions} is nilpotent, \textit{i.e.}, $z_{\text{S}}^{j} = (\alpha \alpha^*)^j = 0$ for all $j>1$, immediately implies
\begin{equation}
    \int \mathcal{D} \alpha \, f (z) = \int \mathcal{D} \alpha \, \left[ f(z_{\text{B}}) + f' (z_{\text{B}}) \alpha \alpha^* \right] = f' (z_{\text{B}}).
    \label{eq:ConcaveAverageIdentity}
\end{equation}
Since $f$ is concave, its first derivative is a monotonically decreasing function, \textit{i.e.}, 
\begin{equation}
    t_1 \le t_2 \Rightarrow f' (t_1) \ge f' (t_2),
\end{equation}
for all $t_1,t_2 \in \mathcal{I}$. Hence, if two phase-space distributions $z_1(\alpha)$ and  $z_2(\alpha)$ satisfy $z_{1,\text{B}}\le z_{2,\text{B}}$, then, using~\eqref{eq:MajorizationRelation} and~\eqref{eq:ConcaveAverageIdentity}, the fermionic phase-space majorization relation $z_1 \prec z_2$ must hold. Thus, majorization relations are ultimately a consequence of $z_{\text{B}}$ being bounded. More precisely, $-1 \le P_{\text{B}} \le 0$ implies $f' (-1) \ge f' (P_{\text{B}}) \ge f' (0)$, from which~\eqref{eq:MajorizationGlauberP} follows, while $-1/2 \le W^-_{\text{B}} \le  0 \le W^+_{\text{B}} \le 1/2$ implies $f' (-1/2) \ge f' (W^-_{\text{B}}) \ge f' (0)\ge f' (W^+_{\text{B}}) \ge f' (1/2)$, which in turn  implies~\eqref{eq:MajorizationWignerW}. Analogously, $0 \le Q_{\text{B}} \le 1$ implies $f' (0) \ge f' (Q_{\text{B}}) \ge f' (1)$, leading to \eqref{eq:MajorizationHusimiQ}. See also~\cite{SM}\textbf{(e)}.

\textit{Second-moment uncertainty relations}---We consider now the fermionic covariance matrix associated with the phase-space distribution $z(\alpha)$, which we define as~\cite{Hackl2021}
\begin{equation}
    \gamma_{j j'} (z) \equiv \frac{1}{2i} \int \mathcal{D} \alpha \, z (\alpha) \, [ \xi_j , \xi_{j'}] = z_{\text{B}} \, (\sigma_y)_{j j'},
\end{equation}
with $j,j'=1,2$ (we set $\xi_1\equiv\alpha$ and $\xi_2\equiv\alpha^*$) and $\sigma_y$ being the second Pauli matrix. For $z=W$, this definition is equivalent to the standard definition of the state's covariance, see~\cite{SM}\textbf{(f)}. Its determinant 
\begin{equation}
    \det \gamma (z) = - z_{\text{B}}^2
\end{equation}
is non-trivially bounded from below by the uncertainty principle (just as for a bosonic mode). Indeed, the determinants of the three covariance matrices read
\begin{equation}
    \begin{split}
        \det \gamma (P) &= - \braket{n}^2, \\
        \det \gamma (W) &= -(1/2 - \braket{n})^2, \\ 
        \det \gamma (Q) &= -(1 - \braket{n})^2,
    \end{split}
\end{equation}
so we have the second-moment uncertainty relations
\begin{equation}
    \begin{split}
        \det \gamma (P_0) &\ge \det \gamma (P) \ge \det \gamma (P_1) = -1, \\
        \det \gamma (W_{1/2}) &\ge \det \gamma (W) \ge \det \gamma (W_{\{0,1\}}) = - 1/4, \\
        \det \gamma (Q_1) &\ge \det \gamma (Q) \ge \det \gamma (Q_0) = - 1,
    \end{split}
    \label{eq:SMURs}
\end{equation}
respectively. The three determinants and their bounds are shown in Fig.~\hyperref[fig:UncertaintyRelations]{2(a)}. Interestingly, $\det \gamma (W) \ge - 1/4$ and $\det \gamma (Q) \ge - 1$ resemble the Robertson-Schrödinger relation~\cite{Robertson1929,Robertson1930,Schroedinger1930} and the uncertainty relation presented in Eq. (73) of~\cite{Haas2021b}, respectively, up to an overall minus sign. 

\textit{Entropic uncertainty relations}---We define the Rényi entropy of a fermionic phase-space distribution $z(\alpha)$ with possibly negative body $z_{\text{B}}$ as
\begin{equation}
    \begin{split}
        S_r (z) &\equiv \frac{1}{1-r} \ln \int \mathcal{D} \alpha \, \abs{z(\alpha)}^r ,
    \end{split}
    \label{eq:RenyiEntropy}
\end{equation}
where $r \in (0,1) \cup (1,\infty)$ denotes the entropic order. The Shannon entropy $S(z)\equiv - \int \mathcal{D} \alpha \, z(\alpha) \ln z(\alpha)$ is recovered in the limit $r \to 1$. For $z_\text{B} \ge 0$,~\eqref{eq:RenyiEntropy} reduces to the standard definition, while for $z_{\text{B}} < 0$, $S_r(z)$ is basically defined as the standard Rényi entropy of $-z>0$. Importantly, the entropies~\eqref{eq:RenyiEntropy} are convex over distributions of the same sign, a striking difference to the bosonic case. 

Following~\eqref{eq:ConcaveAverageIdentity} (or by using the nilpotency of $z_{\text{S}}$), we have that $\int \mathcal{D} \alpha \, \abs{z(\alpha)}^r  = r \abs{z_{\text{B}}}^{r-1} $, which allows to evaluate the Rényi entropy of the three phase-space distributions explicitly as
\begin{equation}
    \begin{split}
        S_r (P) &= \frac{\ln r}{1 - r} - \ln \braket{n}, \\
        S_r (W) &= \frac{\ln r}{1-r} - \ln \big \vert 1/2 - \braket{n} \big \vert, \\
        S_r (Q) &= \frac{\ln r}{1-r} - \ln \left(1 - \braket{n} \right).
    \end{split}
    \label{eq:RenyiEntropiesPhaseSpace}
\end{equation}
Therefore, the entropic uncertainty relations in fermionic phase space are written as
\begin{equation}
    \begin{split}
        S_r (P) &\ge S_r (P_1) = \frac{\ln r}{1-r}, \\
        S_r (W) &\ge S_r (W_{\{0,1\}}) = \frac{\ln r}{1-r} + \ln 2, \\
        S_r (Q) &\ge S_r (Q_0) = \frac{\ln r}{1-r}.
    \end{split}
    \label{eq:EURs}
\end{equation}
We analyze these entropies and their bounds in Fig.~\hyperref[fig:UncertaintyRelations]{2(b)} and  Fig.~\hyperref[fig:UncertaintyRelations]{2(c)}. When taking the limit $r \to 1$, the two uncertainty relations $S(W)\ge -1+\ln 2$ and $S(Q)\ge -1$ resemble the bosonic Wigner entropy conjecture for Wigner-positive states~\cite{VanHerstraeten2021a} and the Wehrl-Lieb inequality~\cite{Wehrl1978,Wehrl1979,Lieb1978}, respectively, except for a minus sign. See~\cite{SM}\textbf{(g)} for details.

\textit{Discussion}---We have derived fermionic uncertainty relations [Eqs.~\eqref{eq:SMURs} and \eqref{eq:EURs}], as well as fermionic majorization relations [Eqs.~\eqref{eq:MajorizationGlauberP}-\eqref{eq:MajorizationHusimiQ}]. These relations are fundamentally implied by Pauli's principle, which states that the vacuum (excited) state has the lowest (highest) occupation. Since all physical states of a single fermionic mode are Gaussian, the moment-based, entropy-based, and majorization-based uncertainty relations are all equivalent (for a given phase-space distribution $z$). First, it is simple to see that $S_r (z) = \ln r /(1-r) - (1/2) \ln [- \det \gamma (z)]$ for all Gaussian $z$; hence~\eqref{eq:SMURs} and~\eqref{eq:EURs} are equivalent. Second, the equivalence between~\eqref{eq:SMURs} and~\eqref{eq:MajorizationGlauberP}-\eqref{eq:MajorizationHusimiQ} is a consequence of how the majorization relation~\eqref{eq:MajorizationRelation} behaves under a change of Grassmann-valued variables, see~\cite{SM}\textbf{(e)}. The Gaussian nature of $z$ is the key feature that enables our exact derivation here, in sharp contrast to bosonic uncertainty relations and majorization relations, which remain challenging problems.

With our analysis, we have paved the ground for studying quantum information problems in fermionic systems. In~\cite{App}, we discuss the implications of our quantum informational approach to fermionic phase-space on quantum cloning and quantum fermionic channels: We demonstrate that a single fermion can be cloned by showing that the no-cloning uncertainty relations do not put any constraints on the clones and discuss how the (in the single-mode case most general) thermal loss channel--describing Gaussian diffusion or a heat flow--increases disorder in phase space. We envision further application in \textit{e.g.}, the context of entanglement theory and quantum cryptography.

Although the fermionic phase-space distributions are themselves Grassmann-valued, all measures of disorder, such as the entropy, are real-valued and, even more importantly, measurable as they can be computed from the occupation numbers of the system. On the theoretical side, it is of particular interest to extend our fermionic uncertainty relations to arbitrary many modes~\cite{Haas2024}, where the set of physical states also contains non-Gaussian states. The proposed methods herein are directly applicable to the two-mode case, where the Wigner majorization relations single out pure Gaussian states as the states of least disorder in phase space. Further, all entropic uncertainty relations acquire an additional factor of two, thereby suggesting an extensive scaling of entropic bounds (just as in the bosonic case)~\cite{TwoFermions}.

\textit{Acknowledgements}---We thank Zacharie Van Herstraeten for valuable discussions on the subject. We acknowledge support from the European Union under project ShoQC within the ERA-NET Cofund in Quantum Technologies (QuantERA) program, as well as from the F.R.S.-FNRS under project CHEQS within the Excellence of Science (EOS) program.


\bibliography{references.bib}

\begin{thebibliography}{76}%
\makeatletter
\providecommand \@ifxundefined [1]{%
 \@ifx{#1\undefined}
}%
\providecommand \@ifnum [1]{%
 \ifnum #1\expandafter \@firstoftwo
 \else \expandafter \@secondoftwo
 \fi
}%
\providecommand \@ifx [1]{%
 \ifx #1\expandafter \@firstoftwo
 \else \expandafter \@secondoftwo
 \fi
}%
\providecommand \natexlab [1]{#1}%
\providecommand \enquote  [1]{``#1''}%
\providecommand \bibnamefont  [1]{#1}%
\providecommand \bibfnamefont [1]{#1}%
\providecommand \citenamefont [1]{#1}%
\providecommand \href@noop [0]{\@secondoftwo}%
\providecommand \href [0]{\begingroup \@sanitize@url \@href}%
\providecommand \@href[1]{\@@startlink{#1}\@@href}%
\providecommand \@@href[1]{\endgroup#1\@@endlink}%
\providecommand \@sanitize@url [0]{\catcode `\\12\catcode `\$12\catcode
  `\&12\catcode `\#12\catcode `\^12\catcode `\_12\catcode `\%12\relax}%
\providecommand \@@startlink[1]{}%
\providecommand \@@endlink[0]{}%
\providecommand \url  [0]{\begingroup\@sanitize@url \@url }%
\providecommand \@url [1]{\endgroup\@href {#1}{\urlprefix }}%
\providecommand \urlprefix  [0]{URL }%
\providecommand \Eprint [0]{\href }%
\providecommand \doibase [0]{https://doi.org/}%
\providecommand \selectlanguage [0]{\@gobble}%
\providecommand \bibinfo  [0]{\@secondoftwo}%
\providecommand \bibfield  [0]{\@secondoftwo}%
\providecommand \translation [1]{[#1]}%
\providecommand \BibitemOpen [0]{}%
\providecommand \bibitemStop [0]{}%
\providecommand \bibitemNoStop [0]{.\EOS\space}%
\providecommand \EOS [0]{\spacefactor3000\relax}%
\providecommand \BibitemShut  [1]{\csname bibitem#1\endcsname}%
\let\auto@bib@innerbib\@empty
\bibitem [{\citenamefont {Heisenberg}(1927)}]{Heisenberg1927}%
  \BibitemOpen
  \bibfield  {author} {\bibinfo {author} {\bibfnamefont {W.}~\bibnamefont
  {Heisenberg}},\ }\bibfield  {title} {\bibinfo {title} {{{\"U}ber den
  anschaulichen Inhalt der quantentheoretischen Kinematik und Mechanik}},\
  }\href {https://doi.org/10.1007/BF01397280} {\bibfield  {journal} {\bibinfo
  {journal} {Z. Phys.}\ }\textbf {\bibinfo {volume} {43}},\ \bibinfo {pages}
  {172} (\bibinfo {year} {1927})}\BibitemShut {NoStop}%
\bibitem [{\citenamefont {Kennard}(1927)}]{Kennard1927}%
  \BibitemOpen
  \bibfield  {author} {\bibinfo {author} {\bibfnamefont {E.~H.}\ \bibnamefont
  {Kennard}},\ }\bibfield  {title} {\bibinfo {title} {{Zur Quantenmechanik
  einfacher Bewegungs-typen}},\ }\href {https://doi.org/10.1007/BF01391200}
  {\bibfield  {journal} {\bibinfo  {journal} {Z. Phys.}\ }\textbf {\bibinfo
  {volume} {44}},\ \bibinfo {pages} {326} (\bibinfo {year} {1927})}\BibitemShut
  {NoStop}%
\bibitem [{\citenamefont {Robertson}(1929)}]{Robertson1929}%
  \BibitemOpen
  \bibfield  {author} {\bibinfo {author} {\bibfnamefont {H.~P.}\ \bibnamefont
  {Robertson}},\ }\bibfield  {title} {\bibinfo {title} {{The Uncertainty
  Principle}},\ }\href {https://doi.org/10.1103/PhysRev.34.163} {\bibfield
  {journal} {\bibinfo  {journal} {Phys. Rev.}\ }\textbf {\bibinfo {volume}
  {34}},\ \bibinfo {pages} {163} (\bibinfo {year} {1929})}\BibitemShut
  {NoStop}%
\bibitem [{\citenamefont {Robertson}(1930)}]{Robertson1930}%
  \BibitemOpen
  \bibfield  {author} {\bibinfo {author} {\bibfnamefont {H.~P.}\ \bibnamefont
  {Robertson}},\ }\bibfield  {title} {\bibinfo {title} {{A general formulation
  of the uncertainty principle and its classical interpretation}},\ }\href
  {https://doi.org/10.1103/PhysRev.35.656} {\bibfield  {journal} {\bibinfo
  {journal} {Phys. Rev.}\ }\textbf {\bibinfo {volume} {35}},\ \bibinfo {pages}
  {656} (\bibinfo {year} {1930})}\BibitemShut {NoStop}%
\bibitem [{\citenamefont {Schrödinger}(1930)}]{Schroedinger1930}%
  \BibitemOpen
  \bibfield  {author} {\bibinfo {author} {\bibfnamefont {E.}~\bibnamefont
  {Schrödinger}},\ }\bibfield  {title} {\bibinfo {title} {{Zum Heisenbergschen
  Unschärfeprinzip}},\ }\href@noop {} {\bibfield  {journal} {\bibinfo
  {journal} {Sitzungsberichte der Preußischen Akademie der Wissenschaften.
  Physikalisch-mathematische Klasse}\ }\textbf {\bibinfo {volume} {14}},\
  \bibinfo {pages} {296} (\bibinfo {year} {1930})}\BibitemShut {NoStop}%
\bibitem [{\citenamefont {Everett}(1957)}]{Everett1957}%
  \BibitemOpen
  \bibfield  {author} {\bibinfo {author} {\bibfnamefont {H.}~\bibnamefont
  {Everett}},\ }\bibfield  {title} {\bibinfo {title} {{'Relative State'
  Formulation of Quantum Mechanics}},\ }\href
  {https://doi.org/10.1103/RevModPhys.29.454} {\bibfield  {journal} {\bibinfo
  {journal} {Rev. Mod. Phys.}\ }\textbf {\bibinfo {volume} {29}},\ \bibinfo
  {pages} {454} (\bibinfo {year} {1957})}\BibitemShut {NoStop}%
\bibitem [{\citenamefont {Beckner}(1975)}]{Beckner1975}%
  \BibitemOpen
  \bibfield  {author} {\bibinfo {author} {\bibfnamefont {W.}~\bibnamefont
  {Beckner}},\ }\bibfield  {title} {\bibinfo {title} {{Inequalities in Fourier
  Analysis}},\ }\href {https://doi.org/10.2307/1970980} {\bibfield  {journal}
  {\bibinfo  {journal} {Ann. Math.}\ }\textbf {\bibinfo {volume} {102}},\
  \bibinfo {pages} {159} (\bibinfo {year} {1975})}\BibitemShut {NoStop}%
\bibitem [{\citenamefont {Deutsch}(1983)}]{Deutsch1983}%
  \BibitemOpen
  \bibfield  {author} {\bibinfo {author} {\bibfnamefont {D.}~\bibnamefont
  {Deutsch}},\ }\bibfield  {title} {\bibinfo {title} {{Uncertainty in Quantum
  Measurements}},\ }\href {https://doi.org/10.1103/PhysRevLett.50.631}
  {\bibfield  {journal} {\bibinfo  {journal} {Phys. Rev. Lett.}\ }\textbf
  {\bibinfo {volume} {50}},\ \bibinfo {pages} {631} (\bibinfo {year}
  {1983})}\BibitemShut {NoStop}%
\bibitem [{\citenamefont {Kraus}(1987)}]{Kraus1987}%
  \BibitemOpen
  \bibfield  {author} {\bibinfo {author} {\bibfnamefont {K.}~\bibnamefont
  {Kraus}},\ }\bibfield  {title} {\bibinfo {title} {{Complementary observables
  and uncertainty relations}},\ }\href
  {https://doi.org/10.1103/PhysRevD.35.3070} {\bibfield  {journal} {\bibinfo
  {journal} {Phys. Rev. D}\ }\textbf {\bibinfo {volume} {35}},\ \bibinfo
  {pages} {3070} (\bibinfo {year} {1987})}\BibitemShut {NoStop}%
\bibitem [{\citenamefont {Maassen}\ and\ \citenamefont
  {Uffink}(1988)}]{Maassen1988}%
  \BibitemOpen
  \bibfield  {author} {\bibinfo {author} {\bibfnamefont {H.}~\bibnamefont
  {Maassen}}\ and\ \bibinfo {author} {\bibfnamefont {J.~B.~M.}\ \bibnamefont
  {Uffink}},\ }\bibfield  {title} {\bibinfo {title} {{Generalized entropic
  uncertainty relations}},\ }\href
  {https://doi.org/10.1103/PhysRevLett.60.1103} {\bibfield  {journal} {\bibinfo
   {journal} {Phys. Rev. Lett.}\ }\textbf {\bibinfo {volume} {60}},\ \bibinfo
  {pages} {1103} (\bibinfo {year} {1988})}\BibitemShut {NoStop}%
\bibitem [{\citenamefont {Berta}\ \emph {et~al.}(2010)\citenamefont {Berta},
  \citenamefont {Christandl}, \citenamefont {Colbeck}, \citenamefont {Renes},\
  and\ \citenamefont {Renner}}]{Berta2010}%
  \BibitemOpen
  \bibfield  {author} {\bibinfo {author} {\bibfnamefont {M.}~\bibnamefont
  {Berta}}, \bibinfo {author} {\bibfnamefont {M.}~\bibnamefont {Christandl}},
  \bibinfo {author} {\bibfnamefont {R.}~\bibnamefont {Colbeck}}, \bibinfo
  {author} {\bibfnamefont {J.~M.}\ \bibnamefont {Renes}},\ and\ \bibinfo
  {author} {\bibfnamefont {R.}~\bibnamefont {Renner}},\ }\bibfield  {title}
  {\bibinfo {title} {{The uncertainty principle in the presence of quantum
  memory}},\ }\href {https://doi.org/10.1038/nphys1734} {\bibfield  {journal}
  {\bibinfo  {journal} {Nat. Phys.}\ }\textbf {\bibinfo {volume} {6}},\
  \bibinfo {pages} {659} (\bibinfo {year} {2010})}\BibitemShut {NoStop}%
\bibitem [{\citenamefont {Wehner}\ and\ \citenamefont
  {Winter}(2010)}]{Wehner2010}%
  \BibitemOpen
  \bibfield  {author} {\bibinfo {author} {\bibfnamefont {S.}~\bibnamefont
  {Wehner}}\ and\ \bibinfo {author} {\bibfnamefont {A.}~\bibnamefont
  {Winter}},\ }\bibfield  {title} {\bibinfo {title} {{Entropic uncertainty
  relations—a survey}},\ }\href
  {https://doi.org/10.1088/1367-2630/12/2/025009} {\bibfield  {journal}
  {\bibinfo  {journal} {New J. Phys.}\ }\textbf {\bibinfo {volume} {12}},\
  \bibinfo {pages} {025009} (\bibinfo {year} {2010})}\BibitemShut {NoStop}%
\bibitem [{\citenamefont {Coles}\ \emph {et~al.}(2012)\citenamefont {Coles},
  \citenamefont {Colbeck}, \citenamefont {Yu},\ and\ \citenamefont
  {Zwolak}}]{Coles2012}%
  \BibitemOpen
  \bibfield  {author} {\bibinfo {author} {\bibfnamefont {P.~J.}\ \bibnamefont
  {Coles}}, \bibinfo {author} {\bibfnamefont {R.}~\bibnamefont {Colbeck}},
  \bibinfo {author} {\bibfnamefont {L.}~\bibnamefont {Yu}},\ and\ \bibinfo
  {author} {\bibfnamefont {M.}~\bibnamefont {Zwolak}},\ }\bibfield  {title}
  {\bibinfo {title} {{Uncertainty Relations from Simple Entropic Properties}},\
  }\href {https://doi.org/10.1103/PhysRevLett.108.210405} {\bibfield  {journal}
  {\bibinfo  {journal} {Phys. Rev. Lett.}\ }\textbf {\bibinfo {volume} {108}},\
  \bibinfo {pages} {210405} (\bibinfo {year} {2012})}\BibitemShut {NoStop}%
\bibitem [{\citenamefont {Bia{\l}ynicki-Birula}\ and\ \citenamefont
  {Mycielski}(1975)}]{Bialynicki-Birula1975}%
  \BibitemOpen
  \bibfield  {author} {\bibinfo {author} {\bibfnamefont {I.}~\bibnamefont
  {Bia{\l}ynicki-Birula}}\ and\ \bibinfo {author} {\bibfnamefont
  {J.}~\bibnamefont {Mycielski}},\ }\bibfield  {title} {\bibinfo {title}
  {{Uncertainty relations for information entropy in wave mechanics}},\ }\href
  {https://doi.org/10.1007/BF01608825} {\bibfield  {journal} {\bibinfo
  {journal} {Commun. Math. Phys.}\ }\textbf {\bibinfo {volume} {44}},\ \bibinfo
  {pages} {129} (\bibinfo {year} {1975})}\BibitemShut {NoStop}%
\bibitem [{\citenamefont {Frank}\ and\ \citenamefont {Lieb}(2012)}]{Frank2012}%
  \BibitemOpen
  \bibfield  {author} {\bibinfo {author} {\bibfnamefont {R.~L.}\ \bibnamefont
  {Frank}}\ and\ \bibinfo {author} {\bibfnamefont {E.~H.}\ \bibnamefont
  {Lieb}},\ }\bibfield  {title} {\bibinfo {title} {{Entropy and the Uncertainty
  Principle}},\ }\href {https://doi.org/10.1007/s00023-012-0175-y} {\bibfield
  {journal} {\bibinfo  {journal} {Ann. Henri Poincaré}\ }\textbf {\bibinfo
  {volume} {13}},\ \bibinfo {pages} {1711} (\bibinfo {year}
  {2012})}\BibitemShut {NoStop}%
\bibitem [{\citenamefont {Hertz}\ and\ \citenamefont {Cerf}(2019)}]{Hertz2019}%
  \BibitemOpen
  \bibfield  {author} {\bibinfo {author} {\bibfnamefont {A.}~\bibnamefont
  {Hertz}}\ and\ \bibinfo {author} {\bibfnamefont {N.~J.}\ \bibnamefont
  {Cerf}},\ }\bibfield  {title} {\bibinfo {title} {Continuous-variable entropic
  uncertainty relations},\ }\href {https://doi.org/10.1088/1751-8121/ab03f3}
  {\bibfield  {journal} {\bibinfo  {journal} {J. Phys. A Math. Theor.}\
  }\textbf {\bibinfo {volume} {52}},\ \bibinfo {pages} {173001} (\bibinfo
  {year} {2019})}\BibitemShut {NoStop}%
\bibitem [{\citenamefont {Van~Herstraeten}\ and\ \citenamefont
  {Cerf}(2021)}]{VanHerstraeten2021a}%
  \BibitemOpen
  \bibfield  {author} {\bibinfo {author} {\bibfnamefont {Z.}~\bibnamefont
  {Van~Herstraeten}}\ and\ \bibinfo {author} {\bibfnamefont {N.~J.}\
  \bibnamefont {Cerf}},\ }\bibfield  {title} {\bibinfo {title} {{Quantum Wigner
  entropy}},\ }\href {https://doi.org/10.1103/PhysRevA.104.042211} {\bibfield
  {journal} {\bibinfo  {journal} {Phys. Rev. A}\ }\textbf {\bibinfo {volume}
  {104}},\ \bibinfo {pages} {042211} (\bibinfo {year} {2021})}\BibitemShut
  {NoStop}%
\bibitem [{\citenamefont {Floerchinger}\ \emph
  {et~al.}(2021{\natexlab{a}})\citenamefont {Floerchinger}, \citenamefont
  {Haas},\ and\ \citenamefont {Hoeber}}]{Haas2021a}%
  \BibitemOpen
  \bibfield  {author} {\bibinfo {author} {\bibfnamefont {S.}~\bibnamefont
  {Floerchinger}}, \bibinfo {author} {\bibfnamefont {T.}~\bibnamefont {Haas}},\
  and\ \bibinfo {author} {\bibfnamefont {B.}~\bibnamefont {Hoeber}},\
  }\bibfield  {title} {\bibinfo {title} {{Relative entropic uncertainty
  relation}},\ }\href {https://doi.org/10.1103/PhysRevA.103.062209} {\bibfield
  {journal} {\bibinfo  {journal} {Phys. Rev. A}\ }\textbf {\bibinfo {volume}
  {103}},\ \bibinfo {pages} {062209} (\bibinfo {year}
  {2021}{\natexlab{a}})}\BibitemShut {NoStop}%
\bibitem [{\citenamefont {Floerchinger}\ \emph
  {et~al.}(2022{\natexlab{a}})\citenamefont {Floerchinger}, \citenamefont
  {Haas},\ and\ \citenamefont {Schröfl}}]{Haas2022b}%
  \BibitemOpen
  \bibfield  {author} {\bibinfo {author} {\bibfnamefont {S.}~\bibnamefont
  {Floerchinger}}, \bibinfo {author} {\bibfnamefont {T.}~\bibnamefont {Haas}},\
  and\ \bibinfo {author} {\bibfnamefont {M.}~\bibnamefont {Schröfl}},\
  }\bibfield  {title} {\bibinfo {title} {{Relative entropic uncertainty
  relation for scalar quantum fields}},\ }\href
  {https://doi.org/10.21468/SciPostPhys.12.3.089} {\bibfield  {journal}
  {\bibinfo  {journal} {SciPost Phys.}\ }\textbf {\bibinfo {volume} {12}},\
  \bibinfo {pages} {089} (\bibinfo {year} {2022}{\natexlab{a}})}\BibitemShut
  {NoStop}%
\bibitem [{\citenamefont {Ditsch}\ and\ \citenamefont {Haas}(2024)}]{Haas2024}%
  \BibitemOpen
  \bibfield  {author} {\bibinfo {author} {\bibfnamefont {S.}~\bibnamefont
  {Ditsch}}\ and\ \bibinfo {author} {\bibfnamefont {T.}~\bibnamefont {Haas}},\
  }\bibfield  {title} {\bibinfo {title} {{Entropic distinguishability of
  quantum fields in phase space}},\ }\href
  {https://doi.org/10.22331/q-2024-07-17-1414} {\bibfield  {journal} {\bibinfo
  {journal} {Quantum}\ }\textbf {\bibinfo {volume} {8}},\ \bibinfo {pages}
  {1414} (\bibinfo {year} {2024})}\BibitemShut {NoStop}%
\bibitem [{\citenamefont {Walborn}\ \emph {et~al.}(2009)\citenamefont
  {Walborn}, \citenamefont {Taketani}, \citenamefont {Salles}, \citenamefont
  {Toscano},\ and\ \citenamefont {de~Matos~Filho}}]{Walborn2009}%
  \BibitemOpen
  \bibfield  {author} {\bibinfo {author} {\bibfnamefont {S.~P.}\ \bibnamefont
  {Walborn}}, \bibinfo {author} {\bibfnamefont {B.~G.}\ \bibnamefont
  {Taketani}}, \bibinfo {author} {\bibfnamefont {A.}~\bibnamefont {Salles}},
  \bibinfo {author} {\bibfnamefont {F.}~\bibnamefont {Toscano}},\ and\ \bibinfo
  {author} {\bibfnamefont {R.~L.}\ \bibnamefont {de~Matos~Filho}},\ }\bibfield
  {title} {\bibinfo {title} {{Entropic Entanglement Criteria for Continuous
  Variables}},\ }\href {https://doi.org/10.1103/PhysRevLett.103.160505}
  {\bibfield  {journal} {\bibinfo  {journal} {Phys. Rev. Lett.}\ }\textbf
  {\bibinfo {volume} {103}},\ \bibinfo {pages} {160505} (\bibinfo {year}
  {2009})}\BibitemShut {NoStop}%
\bibitem [{\citenamefont {Saboia}\ \emph {et~al.}(2011)\citenamefont {Saboia},
  \citenamefont {Toscano},\ and\ \citenamefont {Walborn}}]{Saboia2011}%
  \BibitemOpen
  \bibfield  {author} {\bibinfo {author} {\bibfnamefont {A.}~\bibnamefont
  {Saboia}}, \bibinfo {author} {\bibfnamefont {F.}~\bibnamefont {Toscano}},\
  and\ \bibinfo {author} {\bibfnamefont {S.~P.}\ \bibnamefont {Walborn}},\
  }\bibfield  {title} {\bibinfo {title} {{Family of continuous-variable
  entanglement criteria using general entropy functions}},\ }\href
  {https://doi.org/10.1103/PhysRevA.83.032307} {\bibfield  {journal} {\bibinfo
  {journal} {Phys. Rev. A}\ }\textbf {\bibinfo {volume} {83}},\ \bibinfo
  {pages} {032307} (\bibinfo {year} {2011})}\BibitemShut {NoStop}%
\bibitem [{\citenamefont {Schneeloch}\ \emph {et~al.}(2019)\citenamefont
  {Schneeloch}, \citenamefont {Tison}, \citenamefont {Fanto}, \citenamefont
  {Alsing},\ and\ \citenamefont {Howland}}]{Schneeloch2019}%
  \BibitemOpen
  \bibfield  {author} {\bibinfo {author} {\bibfnamefont {J.}~\bibnamefont
  {Schneeloch}}, \bibinfo {author} {\bibfnamefont {C.~C.}\ \bibnamefont
  {Tison}}, \bibinfo {author} {\bibfnamefont {M.~L.}\ \bibnamefont {Fanto}},
  \bibinfo {author} {\bibfnamefont {P.~M.}\ \bibnamefont {Alsing}},\ and\
  \bibinfo {author} {\bibfnamefont {G.~A.}\ \bibnamefont {Howland}},\
  }\bibfield  {title} {\bibinfo {title} {{Quantifying entanglement in a
  68-billion-dimensional quantum state space}},\ }\href
  {https://doi.org/10.1038/s41467-019-10810-z} {\bibfield  {journal} {\bibinfo
  {journal} {Nat. Commun.}\ }\textbf {\bibinfo {volume} {10}},\ \bibinfo
  {pages} {1} (\bibinfo {year} {2019})}\BibitemShut {NoStop}%
\bibitem [{\citenamefont {Floerchinger}\ \emph
  {et~al.}(2021{\natexlab{b}})\citenamefont {Floerchinger}, \citenamefont
  {Haas},\ and\ \citenamefont {M\"uller-Groeling}}]{Haas2021b}%
  \BibitemOpen
  \bibfield  {author} {\bibinfo {author} {\bibfnamefont {S.}~\bibnamefont
  {Floerchinger}}, \bibinfo {author} {\bibfnamefont {T.}~\bibnamefont {Haas}},\
  and\ \bibinfo {author} {\bibfnamefont {H.}~\bibnamefont
  {M\"uller-Groeling}},\ }\bibfield  {title} {\bibinfo {title} {{Wehrl entropy,
  entropic uncertainty relations, and entanglement}},\ }\href
  {https://doi.org/10.1103/PhysRevA.103.062222} {\bibfield  {journal} {\bibinfo
   {journal} {Phys. Rev. A}\ }\textbf {\bibinfo {volume} {103}},\ \bibinfo
  {pages} {062222} (\bibinfo {year} {2021}{\natexlab{b}})}\BibitemShut
  {NoStop}%
\bibitem [{\citenamefont {Floerchinger}\ \emph
  {et~al.}(2022{\natexlab{b}})\citenamefont {Floerchinger}, \citenamefont
  {G\"arttner}, \citenamefont {Haas},\ and\ \citenamefont
  {Stockdale}}]{Haas2022a}%
  \BibitemOpen
  \bibfield  {author} {\bibinfo {author} {\bibfnamefont {S.}~\bibnamefont
  {Floerchinger}}, \bibinfo {author} {\bibfnamefont {M.}~\bibnamefont
  {G\"arttner}}, \bibinfo {author} {\bibfnamefont {T.}~\bibnamefont {Haas}},\
  and\ \bibinfo {author} {\bibfnamefont {O.~R.}\ \bibnamefont {Stockdale}},\
  }\bibfield  {title} {\bibinfo {title} {{Entropic entanglement criteria in
  phase space}},\ }\href {https://doi.org/10.1103/PhysRevA.105.012409}
  {\bibfield  {journal} {\bibinfo  {journal} {Phys. Rev. A}\ }\textbf {\bibinfo
  {volume} {105}},\ \bibinfo {pages} {012409} (\bibinfo {year}
  {2022}{\natexlab{b}})}\BibitemShut {NoStop}%
\bibitem [{\citenamefont {Grosshans}\ and\ \citenamefont
  {Cerf}(2004)}]{Grosshans2004}%
  \BibitemOpen
  \bibfield  {author} {\bibinfo {author} {\bibfnamefont {F.}~\bibnamefont
  {Grosshans}}\ and\ \bibinfo {author} {\bibfnamefont {N.~J.}\ \bibnamefont
  {Cerf}},\ }\bibfield  {title} {\bibinfo {title} {{Continuous-Variable Quantum
  Cryptography is Secure against Non-Gaussian Attacks}},\ }\href
  {https://doi.org/10.1103/PhysRevLett.92.047905} {\bibfield  {journal}
  {\bibinfo  {journal} {Phys. Rev. Lett.}\ }\textbf {\bibinfo {volume} {92}},\
  \bibinfo {pages} {047905} (\bibinfo {year} {2004})}\BibitemShut {NoStop}%
\bibitem [{\citenamefont {Renes}\ and\ \citenamefont
  {Boileau}(2009)}]{Renes2009}%
  \BibitemOpen
  \bibfield  {author} {\bibinfo {author} {\bibfnamefont {J.~M.}\ \bibnamefont
  {Renes}}\ and\ \bibinfo {author} {\bibfnamefont {J.-C.}\ \bibnamefont
  {Boileau}},\ }\bibfield  {title} {\bibinfo {title} {{Conjectured Strong
  Complementary Information Tradeoff}},\ }\href
  {https://doi.org/10.1103/PhysRevLett.103.020402} {\bibfield  {journal}
  {\bibinfo  {journal} {Phys. Rev. Lett.}\ }\textbf {\bibinfo {volume} {103}},\
  \bibinfo {pages} {020402} (\bibinfo {year} {2009})}\BibitemShut {NoStop}%
\bibitem [{\citenamefont {Tomamichel}\ and\ \citenamefont
  {Renner}(2011)}]{Tomamichel2011}%
  \BibitemOpen
  \bibfield  {author} {\bibinfo {author} {\bibfnamefont {M.}~\bibnamefont
  {Tomamichel}}\ and\ \bibinfo {author} {\bibfnamefont {R.}~\bibnamefont
  {Renner}},\ }\bibfield  {title} {\bibinfo {title} {{Uncertainty Relation for
  Smooth Entropies}},\ }\href {https://doi.org/10.1103/PhysRevLett.106.110506}
  {\bibfield  {journal} {\bibinfo  {journal} {Phys. Rev. Lett.}\ }\textbf
  {\bibinfo {volume} {106}},\ \bibinfo {pages} {110506} (\bibinfo {year}
  {2011})}\BibitemShut {NoStop}%
\bibitem [{\citenamefont {Furrer}\ \emph {et~al.}(2012)\citenamefont {Furrer},
  \citenamefont {Franz}, \citenamefont {Berta}, \citenamefont {Leverrier},
  \citenamefont {Scholz}, \citenamefont {Tomamichel},\ and\ \citenamefont
  {Werner}}]{Furrer2012}%
  \BibitemOpen
  \bibfield  {author} {\bibinfo {author} {\bibfnamefont {F.}~\bibnamefont
  {Furrer}}, \bibinfo {author} {\bibfnamefont {T.}~\bibnamefont {Franz}},
  \bibinfo {author} {\bibfnamefont {M.}~\bibnamefont {Berta}}, \bibinfo
  {author} {\bibfnamefont {A.}~\bibnamefont {Leverrier}}, \bibinfo {author}
  {\bibfnamefont {V.~B.}\ \bibnamefont {Scholz}}, \bibinfo {author}
  {\bibfnamefont {M.}~\bibnamefont {Tomamichel}},\ and\ \bibinfo {author}
  {\bibfnamefont {R.~F.}\ \bibnamefont {Werner}},\ }\bibfield  {title}
  {\bibinfo {title} {{Continuous Variable Quantum Key Distribution: Finite-Key
  Analysis of Composable Security against Coherent Attacks}},\ }\href
  {https://doi.org/10.1103/PhysRevLett.109.100502} {\bibfield  {journal}
  {\bibinfo  {journal} {Phys. Rev. Lett.}\ }\textbf {\bibinfo {volume} {109}},\
  \bibinfo {pages} {100502} (\bibinfo {year} {2012})}\BibitemShut {NoStop}%
\bibitem [{\citenamefont {Tomamichel}\ \emph {et~al.}(2012)\citenamefont
  {Tomamichel}, \citenamefont {Lim}, \citenamefont {Gisin},\ and\ \citenamefont
  {Renner}}]{Tomamichel2012}%
  \BibitemOpen
  \bibfield  {author} {\bibinfo {author} {\bibfnamefont {M.}~\bibnamefont
  {Tomamichel}}, \bibinfo {author} {\bibfnamefont {C.~C.~W.}\ \bibnamefont
  {Lim}}, \bibinfo {author} {\bibfnamefont {N.}~\bibnamefont {Gisin}},\ and\
  \bibinfo {author} {\bibfnamefont {R.}~\bibnamefont {Renner}},\ }\bibfield
  {title} {\bibinfo {title} {{Tight finite-key analysis for quantum
  cryptography}},\ }\href {https://doi.org/10.1038/ncomms1631} {\bibfield
  {journal} {\bibinfo  {journal} {Nat. Comm.}\ }\textbf {\bibinfo {volume}
  {3}},\ \bibinfo {pages} {634} (\bibinfo {year} {2012})}\BibitemShut {NoStop}%
\bibitem [{\citenamefont {Partovi}(2011)}]{Partovi2011}%
  \BibitemOpen
  \bibfield  {author} {\bibinfo {author} {\bibfnamefont {M.~H.}\ \bibnamefont
  {Partovi}},\ }\bibfield  {title} {\bibinfo {title} {{M}ajorization
  formulation of uncertainty in quantum mechanics},\ }\href
  {https://doi.org/10.1103/PhysRevA.84.052117} {\bibfield  {journal} {\bibinfo
  {journal} {Phys. Rev. A}\ }\textbf {\bibinfo {volume} {84}},\ \bibinfo
  {pages} {052117} (\bibinfo {year} {2011})}\BibitemShut {NoStop}%
\bibitem [{\citenamefont {Puchała}\ \emph {et~al.}(2013)\citenamefont
  {Puchała}, \citenamefont {Łukasz Rudnicki},\ and\ \citenamefont
  {Życzkowski}}]{Puchala2013}%
  \BibitemOpen
  \bibfield  {author} {\bibinfo {author} {\bibfnamefont {Z.}~\bibnamefont
  {Puchała}}, \bibinfo {author} {\bibnamefont {Łukasz Rudnicki}},\ and\
  \bibinfo {author} {\bibfnamefont {K.}~\bibnamefont {Życzkowski}},\
  }\bibfield  {title} {\bibinfo {title} {{Majorization entropic uncertainty
  relations}},\ }\href {https://doi.org/10.1088/1751-8113/46/27/272002}
  {\bibfield  {journal} {\bibinfo  {journal} {J. Phys. A Math. Theor.}\
  }\textbf {\bibinfo {volume} {46}},\ \bibinfo {pages} {272002} (\bibinfo
  {year} {2013})}\BibitemShut {NoStop}%
\bibitem [{\citenamefont {Narasimhachar}\ \emph {et~al.}(2016)\citenamefont
  {Narasimhachar}, \citenamefont {Poostindouz},\ and\ \citenamefont
  {Gour}}]{Narasimhachar2016}%
  \BibitemOpen
  \bibfield  {author} {\bibinfo {author} {\bibfnamefont {V.}~\bibnamefont
  {Narasimhachar}}, \bibinfo {author} {\bibfnamefont {A.}~\bibnamefont
  {Poostindouz}},\ and\ \bibinfo {author} {\bibfnamefont {G.}~\bibnamefont
  {Gour}},\ }\bibfield  {title} {\bibinfo {title} {{Uncertainty, joint
  uncertainty, and the quantum uncertainty principle}},\ }\href
  {https://doi.org/10.1088/1367-2630/18/3/033019} {\bibfield  {journal}
  {\bibinfo  {journal} {New J. Phys.}\ }\textbf {\bibinfo {volume} {18}},\
  \bibinfo {pages} {033019} (\bibinfo {year} {2016})}\BibitemShut {NoStop}%
\bibitem [{\citenamefont {Lieb}\ and\ \citenamefont
  {Solovej}(2014)}]{Lieb2014b}%
  \BibitemOpen
  \bibfield  {author} {\bibinfo {author} {\bibfnamefont {E.~H.}\ \bibnamefont
  {Lieb}}\ and\ \bibinfo {author} {\bibfnamefont {J.~P.}\ \bibnamefont
  {Solovej}},\ }\bibfield  {title} {\bibinfo {title} {{Proof of an entropy
  conjecture for Bloch coherent spin states and its generalizations}},\ }\href
  {https://doi.org/10.1007/s11511-014-0113-6} {\bibfield  {journal} {\bibinfo
  {journal} {Acta Math.}\ }\textbf {\bibinfo {volume} {212}},\ \bibinfo {pages}
  {379} (\bibinfo {year} {2014})}\BibitemShut {NoStop}%
\bibitem [{\citenamefont {Lieb}\ and\ \citenamefont
  {Solovej}(2016)}]{Lieb2016}%
  \BibitemOpen
  \bibfield  {author} {\bibinfo {author} {\bibfnamefont {E.~H.}\ \bibnamefont
  {Lieb}}\ and\ \bibinfo {author} {\bibfnamefont {J.~P.}\ \bibnamefont
  {Solovej}},\ }\bibfield  {title} {\bibinfo {title} {{Proof of the Wehrl-type
  Entropy Conjecture for Symmetric ${SU(N)}$ Coherent States}},\ }\href
  {https://doi.org/10.1007/s00220-016-2596-9} {\bibfield  {journal} {\bibinfo
  {journal} {Commun. Math. Phys.}\ }\textbf {\bibinfo {volume} {348}},\
  \bibinfo {pages} {567} (\bibinfo {year} {2016})}\BibitemShut {NoStop}%
\bibitem [{\citenamefont {Lieb}\ and\ \citenamefont
  {Solovej}(2021)}]{Lieb2021}%
  \BibitemOpen
  \bibfield  {author} {\bibinfo {author} {\bibfnamefont {E.~H.}\ \bibnamefont
  {Lieb}}\ and\ \bibinfo {author} {\bibfnamefont {J.~P.}\ \bibnamefont
  {Solovej}},\ }\bibinfo {title} {{Wehrl-type coherent state entropy
  inequalities for SU(1,1) and its AX+B subgroup}},\ in\ \href
  {https://doi.org/10.4171/ECR/18-1/18} {\emph {\bibinfo {booktitle} {Partial
  Differential Equations, Spectral Theory, and Mathematical Physics}}}\
  (\bibinfo {year} {2021})\ p.\ \bibinfo {pages} {301–314}\BibitemShut
  {NoStop}%
\bibitem [{\citenamefont {Van~Herstraeten}\ \emph {et~al.}(2023)\citenamefont
  {Van~Herstraeten}, \citenamefont {Jabbour},\ and\ \citenamefont
  {Cerf}}]{VanHerstraeten2021b}%
  \BibitemOpen
  \bibfield  {author} {\bibinfo {author} {\bibfnamefont {Z.}~\bibnamefont
  {Van~Herstraeten}}, \bibinfo {author} {\bibfnamefont {M.~G.}\ \bibnamefont
  {Jabbour}},\ and\ \bibinfo {author} {\bibfnamefont {N.~J.}\ \bibnamefont
  {Cerf}},\ }\bibfield  {title} {\bibinfo {title} {{Continuous majorization in
  quantum phase space}},\ }\href {https://doi.org/10.22331/q-2023-05-24-1021}
  {\bibfield  {journal} {\bibinfo  {journal} {{Quantum}}\ }\textbf {\bibinfo
  {volume} {7}},\ \bibinfo {pages} {1021} (\bibinfo {year} {2023})}\BibitemShut
  {NoStop}%
\bibitem [{\citenamefont {Nielsen}(1999)}]{Nielsen1999}%
  \BibitemOpen
  \bibfield  {author} {\bibinfo {author} {\bibfnamefont {M.~A.}\ \bibnamefont
  {Nielsen}},\ }\bibfield  {title} {\bibinfo {title} {{Conditions for a Class
  of Entanglement Transformations}},\ }\href
  {https://doi.org/10.1103/PhysRevLett.83.436} {\bibfield  {journal} {\bibinfo
  {journal} {Phys. Rev. Lett.}\ }\textbf {\bibinfo {volume} {83}},\ \bibinfo
  {pages} {436} (\bibinfo {year} {1999})}\BibitemShut {NoStop}%
\bibitem [{\citenamefont {G\"arttner}\ \emph
  {et~al.}(2023{\natexlab{a}})\citenamefont {G\"arttner}, \citenamefont
  {Haas},\ and\ \citenamefont {Noll}}]{Haas2022d}%
  \BibitemOpen
  \bibfield  {author} {\bibinfo {author} {\bibfnamefont {M.}~\bibnamefont
  {G\"arttner}}, \bibinfo {author} {\bibfnamefont {T.}~\bibnamefont {Haas}},\
  and\ \bibinfo {author} {\bibfnamefont {J.}~\bibnamefont {Noll}},\ }\bibfield
  {title} {\bibinfo {title} {{General Class of Continuous Variable Entanglement
  Criteria}},\ }\href {https://doi.org/10.1103/PhysRevLett.131.150201}
  {\bibfield  {journal} {\bibinfo  {journal} {Phys. Rev. Lett.}\ }\textbf
  {\bibinfo {volume} {131}},\ \bibinfo {pages} {150201} (\bibinfo {year}
  {2023}{\natexlab{a}})}\BibitemShut {NoStop}%
\bibitem [{\citenamefont {G\"arttner}\ \emph
  {et~al.}(2023{\natexlab{b}})\citenamefont {G\"arttner}, \citenamefont
  {Haas},\ and\ \citenamefont {Noll}}]{Haas2022c}%
  \BibitemOpen
  \bibfield  {author} {\bibinfo {author} {\bibfnamefont {M.}~\bibnamefont
  {G\"arttner}}, \bibinfo {author} {\bibfnamefont {T.}~\bibnamefont {Haas}},\
  and\ \bibinfo {author} {\bibfnamefont {J.}~\bibnamefont {Noll}},\ }\bibfield
  {title} {\bibinfo {title} {{Detecting continuous-variable entanglement in
  phase space with the $Q$ distribution}},\ }\href
  {https://doi.org/10.1103/PhysRevA.108.042410} {\bibfield  {journal} {\bibinfo
   {journal} {Phys. Rev. A}\ }\textbf {\bibinfo {volume} {108}},\ \bibinfo
  {pages} {042410} (\bibinfo {year} {2023}{\natexlab{b}})}\BibitemShut
  {NoStop}%
\bibitem [{\citenamefont {Weedbrook}\ \emph {et~al.}(2012)\citenamefont
  {Weedbrook}, \citenamefont {Pirandola}, \citenamefont
  {Garc\'{i}a-Patr\'{o}n}, \citenamefont {Cerf}, \citenamefont {Ralph},
  \citenamefont {Shapiro},\ and\ \citenamefont {Lloyd}}]{Weedbrook2012}%
  \BibitemOpen
  \bibfield  {author} {\bibinfo {author} {\bibfnamefont {C.}~\bibnamefont
  {Weedbrook}}, \bibinfo {author} {\bibfnamefont {S.}~\bibnamefont
  {Pirandola}}, \bibinfo {author} {\bibfnamefont {R.}~\bibnamefont
  {Garc\'{i}a-Patr\'{o}n}}, \bibinfo {author} {\bibfnamefont {N.~J.}\
  \bibnamefont {Cerf}}, \bibinfo {author} {\bibfnamefont {T.~C.}\ \bibnamefont
  {Ralph}}, \bibinfo {author} {\bibfnamefont {J.~H.}\ \bibnamefont {Shapiro}},\
  and\ \bibinfo {author} {\bibfnamefont {S.}~\bibnamefont {Lloyd}},\ }\bibfield
   {title} {\bibinfo {title} {{Gaussian quantum information}},\ }\href
  {https://doi.org/10.1103/RevModPhys.84.621} {\bibfield  {journal} {\bibinfo
  {journal} {Rev. Mod. Phys.}\ }\textbf {\bibinfo {volume} {84}},\ \bibinfo
  {pages} {621} (\bibinfo {year} {2012})}\BibitemShut {NoStop}%
\bibitem [{\citenamefont {Serafini}(2017)}]{Serafini2017}%
  \BibitemOpen
  \bibfield  {author} {\bibinfo {author} {\bibfnamefont {A.}~\bibnamefont
  {Serafini}},\ }\href {https://doi.org/10.1201/9781315118727} {\emph {\bibinfo
  {title} {{Quantum Continuous Variables}}}}\ (\bibinfo  {publisher} {CRC
  Press},\ \bibinfo {year} {2017})\BibitemShut {NoStop}%
\bibitem [{\citenamefont {Nielsen}\ and\ \citenamefont
  {Chuang}(2010)}]{Nielsen2010}%
  \BibitemOpen
  \bibfield  {author} {\bibinfo {author} {\bibfnamefont {M.~A.}\ \bibnamefont
  {Nielsen}}\ and\ \bibinfo {author} {\bibfnamefont {I.~L.}\ \bibnamefont
  {Chuang}},\ }\href {https://doi.org/10.1017/CBO9780511976667} {\emph
  {\bibinfo {title} {{Quantum Computation and Quantum Information: 10th
  Anniversary Edition}}}}\ (\bibinfo  {publisher} {Cambridge University
  Press},\ \bibinfo {year} {2010})\BibitemShut {NoStop}%
\bibitem [{\citenamefont {Wilde}(2013)}]{Wilde2013}%
  \BibitemOpen
  \bibfield  {author} {\bibinfo {author} {\bibfnamefont {M.~M.}\ \bibnamefont
  {Wilde}},\ }\href {https://doi.org/10.1017/CBO9781139525343} {\emph {\bibinfo
  {title} {{Quantum Information Theory}}}}\ (\bibinfo  {publisher} {Cambridge
  University Press},\ \bibinfo {year} {2013})\BibitemShut {NoStop}%
\bibitem [{\citenamefont {Pauli}(1925)}]{Pauli1925}%
  \BibitemOpen
  \bibfield  {author} {\bibinfo {author} {\bibfnamefont {W.}~\bibnamefont
  {Pauli}},\ }\bibfield  {title} {\bibinfo {title} {{Über den Zusammenhang des
  Abschlusses der Elektronengruppen im Atom mit der Komplexstruktur der
  Spektren}},\ }\href {https://doi.org/10.1007/BF02980631} {\bibfield
  {journal} {\bibinfo  {journal} {Z. Physik}\ }\textbf {\bibinfo {volume}
  {31}},\ \bibinfo {pages} {765} (\bibinfo {year} {1925})}\BibitemShut
  {NoStop}%
\bibitem [{\citenamefont {Cahill}\ and\ \citenamefont
  {Glauber}(1999)}]{Cahill1999}%
  \BibitemOpen
  \bibfield  {author} {\bibinfo {author} {\bibfnamefont {K.~E.}\ \bibnamefont
  {Cahill}}\ and\ \bibinfo {author} {\bibfnamefont {R.~J.}\ \bibnamefont
  {Glauber}},\ }\bibfield  {title} {\bibinfo {title} {{Density operators for
  fermions}},\ }\href {https://doi.org/10.1103/PhysRevA.59.1538} {\bibfield
  {journal} {\bibinfo  {journal} {Phys. Rev. A}\ }\textbf {\bibinfo {volume}
  {59}},\ \bibinfo {pages} {1538} (\bibinfo {year} {1999})}\BibitemShut
  {NoStop}%
\bibitem [{\citenamefont {Hackl}\ and\ \citenamefont
  {Bianchi}(2021)}]{Hackl2021}%
  \BibitemOpen
  \bibfield  {author} {\bibinfo {author} {\bibfnamefont {L.}~\bibnamefont
  {Hackl}}\ and\ \bibinfo {author} {\bibfnamefont {E.}~\bibnamefont
  {Bianchi}},\ }\bibfield  {title} {\bibinfo {title} {{Bosonic and fermionic
  Gaussian states from Kähler structures}},\ }\href
  {https://doi.org/10.21468/SciPostPhysCore.4.3.025} {\bibfield  {journal}
  {\bibinfo  {journal} {SciPost Phys. Core}\ }\textbf {\bibinfo {volume} {4}},\
  \bibinfo {pages} {025} (\bibinfo {year} {2021})}\BibitemShut {NoStop}%
\bibitem [{\citenamefont {Ba\~nuls}\ \emph {et~al.}(2007)\citenamefont
  {Ba\~nuls}, \citenamefont {Cirac},\ and\ \citenamefont {Wolf}}]{Banules2007}%
  \BibitemOpen
  \bibfield  {author} {\bibinfo {author} {\bibfnamefont {M.-C.}\ \bibnamefont
  {Ba\~nuls}}, \bibinfo {author} {\bibfnamefont {J.~I.}\ \bibnamefont
  {Cirac}},\ and\ \bibinfo {author} {\bibfnamefont {M.~M.}\ \bibnamefont
  {Wolf}},\ }\bibfield  {title} {\bibinfo {title} {{Entanglement in fermionic
  systems}},\ }\href {https://doi.org/10.1103/PhysRevA.76.022311} {\bibfield
  {journal} {\bibinfo  {journal} {Phys. Rev. A}\ }\textbf {\bibinfo {volume}
  {76}},\ \bibinfo {pages} {022311} (\bibinfo {year} {2007})}\BibitemShut
  {NoStop}%
\bibitem [{\citenamefont {Zander}\ and\ \citenamefont
  {Plastino}(2010)}]{Zander2010}%
  \BibitemOpen
  \bibfield  {author} {\bibinfo {author} {\bibfnamefont {C.}~\bibnamefont
  {Zander}}\ and\ \bibinfo {author} {\bibfnamefont {A.~R.}\ \bibnamefont
  {Plastino}},\ }\bibfield  {title} {\bibinfo {title} {{Uncertainty relations
  and entanglement in fermion systems}},\ }\href
  {https://doi.org/10.1103/PhysRevA.81.062128} {\bibfield  {journal} {\bibinfo
  {journal} {Phys. Rev. A}\ }\textbf {\bibinfo {volume} {81}},\ \bibinfo
  {pages} {062128} (\bibinfo {year} {2010})}\BibitemShut {NoStop}%
\bibitem [{\citenamefont {Friis}\ \emph {et~al.}(2013)\citenamefont {Friis},
  \citenamefont {Lee},\ and\ \citenamefont {Bruschi}}]{Friis2013}%
  \BibitemOpen
  \bibfield  {author} {\bibinfo {author} {\bibfnamefont {N.}~\bibnamefont
  {Friis}}, \bibinfo {author} {\bibfnamefont {A.~R.}\ \bibnamefont {Lee}},\
  and\ \bibinfo {author} {\bibfnamefont {D.~E.}\ \bibnamefont {Bruschi}},\
  }\bibfield  {title} {\bibinfo {title} {{Fermionic-mode entanglement in
  quantum information}},\ }\href {https://doi.org/10.1103/PhysRevA.87.022338}
  {\bibfield  {journal} {\bibinfo  {journal} {Phys. Rev. A}\ }\textbf {\bibinfo
  {volume} {87}},\ \bibinfo {pages} {022338} (\bibinfo {year}
  {2013})}\BibitemShut {NoStop}%
\bibitem [{\citenamefont {Friis}(2016)}]{Friis2016}%
  \BibitemOpen
  \bibfield  {author} {\bibinfo {author} {\bibfnamefont {N.}~\bibnamefont
  {Friis}},\ }\bibfield  {title} {\bibinfo {title} {{Reasonable fermionic
  quantum information theories require relativity}},\ }\href
  {https://doi.org/10.1088/1367-2630/18/3/033014} {\bibfield  {journal}
  {\bibinfo  {journal} {New J. Phys.}\ }\textbf {\bibinfo {volume} {18}},\
  \bibinfo {pages} {033014} (\bibinfo {year} {2016})}\BibitemShut {NoStop}%
\bibitem [{\citenamefont {Debarba}\ \emph {et~al.}(2020)\citenamefont
  {Debarba}, \citenamefont {Iemini}, \citenamefont {Giedke},\ and\
  \citenamefont {Friis}}]{Debarba2020}%
  \BibitemOpen
  \bibfield  {author} {\bibinfo {author} {\bibfnamefont {T.}~\bibnamefont
  {Debarba}}, \bibinfo {author} {\bibfnamefont {F.}~\bibnamefont {Iemini}},
  \bibinfo {author} {\bibfnamefont {G.}~\bibnamefont {Giedke}},\ and\ \bibinfo
  {author} {\bibfnamefont {N.}~\bibnamefont {Friis}},\ }\bibfield  {title}
  {\bibinfo {title} {{Teleporting quantum information encoded in fermionic
  modes}},\ }\href {https://doi.org/10.1103/PhysRevA.101.052326} {\bibfield
  {journal} {\bibinfo  {journal} {Phys. Rev. A}\ }\textbf {\bibinfo {volume}
  {101}},\ \bibinfo {pages} {052326} (\bibinfo {year} {2020})}\BibitemShut
  {NoStop}%
\bibitem [{\citenamefont {Casini}\ and\ \citenamefont
  {Huerta}(2009)}]{Casini2009}%
  \BibitemOpen
  \bibfield  {author} {\bibinfo {author} {\bibfnamefont {H.}~\bibnamefont
  {Casini}}\ and\ \bibinfo {author} {\bibfnamefont {M.}~\bibnamefont
  {Huerta}},\ }\bibfield  {title} {\bibinfo {title} {Entanglement entropy in
  free quantum field theory},\ }\href
  {https://doi.org/10.1088/1751-8113/42/50/504007} {\bibfield  {journal}
  {\bibinfo  {journal} {J. Phys. A Math. Theo.}\ }\textbf {\bibinfo {volume}
  {42}},\ \bibinfo {pages} {504007} (\bibinfo {year} {2009})}\BibitemShut
  {NoStop}%
\bibitem [{\citenamefont {Calabrese}\ and\ \citenamefont
  {Cardy}(2004)}]{Calabrese2004}%
  \BibitemOpen
  \bibfield  {author} {\bibinfo {author} {\bibfnamefont {P.}~\bibnamefont
  {Calabrese}}\ and\ \bibinfo {author} {\bibfnamefont {J.}~\bibnamefont
  {Cardy}},\ }\bibfield  {title} {\bibinfo {title} {Entanglement entropy and
  quantum field theory},\ }\href
  {https://doi.org/10.1088/1742-5468/2004/06/P06002} {\bibfield  {journal}
  {\bibinfo  {journal} {J. Stat. Mech. Theo. Exp.}\ }\textbf {\bibinfo {volume}
  {2004}},\ \bibinfo {pages} {P06002} (\bibinfo {year} {2004})}\BibitemShut
  {NoStop}%
\bibitem [{\citenamefont {Calabrese}\ and\ \citenamefont
  {Cardy}(2009)}]{Calabrese2009}%
  \BibitemOpen
  \bibfield  {author} {\bibinfo {author} {\bibfnamefont {P.}~\bibnamefont
  {Calabrese}}\ and\ \bibinfo {author} {\bibfnamefont {J.}~\bibnamefont
  {Cardy}},\ }\bibfield  {title} {\bibinfo {title} {Entanglement entropy and
  conformal field theory},\ }\href
  {https://doi.org/10.1088/1751-8113/42/50/504005} {\bibfield  {journal}
  {\bibinfo  {journal} {J. Phys. A Math. Theo.}\ }\textbf {\bibinfo {volume}
  {42}},\ \bibinfo {pages} {504005} (\bibinfo {year} {2009})}\BibitemShut
  {NoStop}%
\bibitem [{SM()}]{SM}%
  \BibitemOpen
  \href@noop {} {}\bibinfo {note} {See Supplemental Material for
  details.}\BibitemShut {Stop}%
\bibitem [{\citenamefont {Hegerfeldt}\ \emph {et~al.}(1968)\citenamefont
  {Hegerfeldt}, \citenamefont {Kraus},\ and\ \citenamefont
  {Wigner}}]{Hegerfeldt1968}%
  \BibitemOpen
  \bibfield  {author} {\bibinfo {author} {\bibfnamefont {G.~C.}\ \bibnamefont
  {Hegerfeldt}}, \bibinfo {author} {\bibfnamefont {K.}~\bibnamefont {Kraus}},\
  and\ \bibinfo {author} {\bibfnamefont {E.~P.}\ \bibnamefont {Wigner}},\
  }\bibfield  {title} {\bibinfo {title} {{Proof of the Fermion Superselection
  Rule without the Assumption of Time‐Reversal Invariance}},\ }\href
  {https://doi.org/10.1063/1.1664539} {\bibfield  {journal} {\bibinfo
  {journal} {J. Math. Phys.}\ }\textbf {\bibinfo {volume} {9}},\ \bibinfo
  {pages} {2029} (\bibinfo {year} {1968})}\BibitemShut {NoStop}%
\bibitem [{\citenamefont {DeWitt}(1992)}]{DeWitt1992}%
  \BibitemOpen
  \bibfield  {author} {\bibinfo {author} {\bibfnamefont {B.}~\bibnamefont
  {DeWitt}},\ }\href {https://doi.org/10.1017/CBO9780511564000} {\emph
  {\bibinfo {title} {{Supermanifolds}}}},\ \bibinfo {edition} {2nd}\ ed.,\
  Cambridge Monographs on Mathematical Physics\ (\bibinfo  {publisher}
  {Cambridge University Press},\ \bibinfo {year} {1992})\BibitemShut {NoStop}%
\bibitem [{\citenamefont {Marshall}\ \emph {et~al.}(2011)\citenamefont
  {Marshall}, \citenamefont {Olkin},\ and\ \citenamefont
  {Arnold}}]{Marshall2011}%
  \BibitemOpen
  \bibfield  {author} {\bibinfo {author} {\bibfnamefont {A.~W.}\ \bibnamefont
  {Marshall}}, \bibinfo {author} {\bibfnamefont {I.}~\bibnamefont {Olkin}},\
  and\ \bibinfo {author} {\bibfnamefont {B.~C.}\ \bibnamefont {Arnold}},\
  }\href {https://doi.org/10.1007/978-0-387-68276-1} {\emph {\bibinfo {title}
  {{Inequalities: Theory of Majorization and Its Applications}}}}\ (\bibinfo
  {publisher} {Springer New York},\ \bibinfo {year} {2011})\BibitemShut
  {NoStop}%
\bibitem [{\citenamefont {Wehrl}(1978)}]{Wehrl1978}%
  \BibitemOpen
  \bibfield  {author} {\bibinfo {author} {\bibfnamefont {A.}~\bibnamefont
  {Wehrl}},\ }\bibfield  {title} {\bibinfo {title} {{General properties of
  entropy}},\ }\href {https://doi.org/10.1103/RevModPhys.50.221} {\bibfield
  {journal} {\bibinfo  {journal} {Rev. Mod. Phys.}\ }\textbf {\bibinfo {volume}
  {50}},\ \bibinfo {pages} {221} (\bibinfo {year} {1978})}\BibitemShut
  {NoStop}%
\bibitem [{\citenamefont {Wehrl}(1979)}]{Wehrl1979}%
  \BibitemOpen
  \bibfield  {author} {\bibinfo {author} {\bibfnamefont {A.}~\bibnamefont
  {Wehrl}},\ }\bibfield  {title} {\bibinfo {title} {{On the relation between
  classical and quantum-mechanical entropy}},\ }\href
  {https://doi.org/10.1016/0034-4877(79)90070-3} {\bibfield  {journal}
  {\bibinfo  {journal} {Rep. Math. Phys.}\ }\textbf {\bibinfo {volume} {16}},\
  \bibinfo {pages} {353} (\bibinfo {year} {1979})}\BibitemShut {NoStop}%
\bibitem [{\citenamefont {Lieb}(1978)}]{Lieb1978}%
  \BibitemOpen
  \bibfield  {author} {\bibinfo {author} {\bibfnamefont {E.~H.}\ \bibnamefont
  {Lieb}},\ }\bibfield  {title} {\bibinfo {title} {{Proof of an entropy
  conjecture of Wehrl}},\ }\href {https://doi.org/10.1007/BF01940328}
  {\bibfield  {journal} {\bibinfo  {journal} {Commun. Math. Phys.}\ }\textbf
  {\bibinfo {volume} {62}},\ \bibinfo {pages} {35} (\bibinfo {year}
  {1978})}\BibitemShut {NoStop}%
\bibitem [{App()}]{App}%
  \BibitemOpen
  \href@noop {} {}\bibinfo {note} {See End Matter for details.}\BibitemShut
  {Stop}%
\bibitem [{\citenamefont {Cerf}\ and\ \citenamefont
  {Haas}(2024)}]{TwoFermions}%
  \BibitemOpen
  \bibfield  {author} {\bibinfo {author} {\bibfnamefont {N.~J.}\ \bibnamefont
  {Cerf}}\ and\ \bibinfo {author} {\bibfnamefont {T.}~\bibnamefont {Haas}},\
  }\href@noop {} {\bibinfo {title} {{Majorization and uncertainty relations for
  multiple fermionic modes}}} (\bibinfo {year} {2024}),\ \bibinfo {note} {in
  preparation}\BibitemShut {NoStop}%
\bibitem [{\citenamefont {Wootters}\ and\ \citenamefont
  {Zurek}(1981)}]{Wootters1982}%
  \BibitemOpen
  \bibfield  {author} {\bibinfo {author} {\bibfnamefont {W.~K.}\ \bibnamefont
  {Wootters}}\ and\ \bibinfo {author} {\bibfnamefont {W.~H.}\ \bibnamefont
  {Zurek}},\ }\bibfield  {title} {\bibinfo {title} {{A single quantum cannot be
  cloned}},\ }\href {https://doi.org/10.1038/299802a0} {\bibfield  {journal}
  {\bibinfo  {journal} {Nature}\ }\textbf {\bibinfo {volume} {299}},\ \bibinfo
  {pages} {802} (\bibinfo {year} {1981})}\BibitemShut {NoStop}%
\bibitem [{\citenamefont {Dieks}(1982)}]{Dieks1982}%
  \BibitemOpen
  \bibfield  {author} {\bibinfo {author} {\bibfnamefont {D.}~\bibnamefont
  {Dieks}},\ }\bibfield  {title} {\bibinfo {title} {{Communication by EPR
  devices}},\ }\href
  {https://doi.org/https://doi.org/10.1016/0375-9601(82)90084-6} {\bibfield
  {journal} {\bibinfo  {journal} {Phys. Lett. A}\ }\textbf {\bibinfo {volume}
  {92}},\ \bibinfo {pages} {271} (\bibinfo {year} {1982})}\BibitemShut
  {NoStop}%
\bibitem [{\citenamefont {Scarani}\ \emph {et~al.}(2005)\citenamefont
  {Scarani}, \citenamefont {Iblisdir}, \citenamefont {Gisin},\ and\
  \citenamefont {Ac\'{\i}n}}]{Scarani2005}%
  \BibitemOpen
  \bibfield  {author} {\bibinfo {author} {\bibfnamefont {V.}~\bibnamefont
  {Scarani}}, \bibinfo {author} {\bibfnamefont {S.}~\bibnamefont {Iblisdir}},
  \bibinfo {author} {\bibfnamefont {N.}~\bibnamefont {Gisin}},\ and\ \bibinfo
  {author} {\bibfnamefont {A.}~\bibnamefont {Ac\'{\i}n}},\ }\bibfield  {title}
  {\bibinfo {title} {{Quantum cloning}},\ }\href
  {https://doi.org/10.1103/RevModPhys.77.1225} {\bibfield  {journal} {\bibinfo
  {journal} {Rev. Mod. Phys.}\ }\textbf {\bibinfo {volume} {77}},\ \bibinfo
  {pages} {1225} (\bibinfo {year} {2005})}\BibitemShut {NoStop}%
\bibitem [{\citenamefont {Bu\ifmmode~\check{z}\else \v{z}\fi{}ek}\ and\
  \citenamefont {Hillery}(1996)}]{Buzek1996}%
  \BibitemOpen
  \bibfield  {author} {\bibinfo {author} {\bibfnamefont {V.}~\bibnamefont
  {Bu\ifmmode~\check{z}\else \v{z}\fi{}ek}}\ and\ \bibinfo {author}
  {\bibfnamefont {M.}~\bibnamefont {Hillery}},\ }\bibfield  {title} {\bibinfo
  {title} {{Quantum copying: Beyond the no-cloning theorem}},\ }\href
  {https://doi.org/10.1103/PhysRevA.54.1844} {\bibfield  {journal} {\bibinfo
  {journal} {Phys. Rev. A}\ }\textbf {\bibinfo {volume} {54}},\ \bibinfo
  {pages} {1844} (\bibinfo {year} {1996})}\BibitemShut {NoStop}%
\bibitem [{\citenamefont {Cerf}\ \emph {et~al.}(2000)\citenamefont {Cerf},
  \citenamefont {Ipe},\ and\ \citenamefont {Rottenberg}}]{Cerf2000}%
  \BibitemOpen
  \bibfield  {author} {\bibinfo {author} {\bibfnamefont {N.~J.}\ \bibnamefont
  {Cerf}}, \bibinfo {author} {\bibfnamefont {A.}~\bibnamefont {Ipe}},\ and\
  \bibinfo {author} {\bibfnamefont {X.}~\bibnamefont {Rottenberg}},\ }\bibfield
   {title} {\bibinfo {title} {{Cloning of Continuous Quantum Variables}},\
  }\href {https://doi.org/10.1103/PhysRevLett.85.1754} {\bibfield  {journal}
  {\bibinfo  {journal} {Phys. Rev. Lett.}\ }\textbf {\bibinfo {volume} {85}},\
  \bibinfo {pages} {1754} (\bibinfo {year} {2000})}\BibitemShut {NoStop}%
\bibitem [{\citenamefont {Cerf}(2003)}]{Cerf2003}%
  \BibitemOpen
  \bibfield  {author} {\bibinfo {author} {\bibfnamefont {N.~J.}\ \bibnamefont
  {Cerf}},\ }\bibinfo {title} {{Quantum Cloning with Continuous Variables}},\
  in\ \href {https://doi.org/10.1007/978-94-015-1258-9_20} {\emph {\bibinfo
  {booktitle} {{Quantum Information with Continuous Variables}}}},\ \bibinfo
  {editor} {edited by\ \bibinfo {editor} {\bibfnamefont {S.~L.}\ \bibnamefont
  {Braunstein}}\ and\ \bibinfo {editor} {\bibfnamefont {A.~K.}\ \bibnamefont
  {Pati}}}\ (\bibinfo  {publisher} {Springer Netherlands},\ \bibinfo {address}
  {Dordrecht},\ \bibinfo {year} {2003})\ pp.\ \bibinfo {pages}
  {277--293}\BibitemShut {NoStop}%
\bibitem [{\citenamefont {Barnum}\ \emph {et~al.}(1996)\citenamefont {Barnum},
  \citenamefont {Caves}, \citenamefont {Fuchs}, \citenamefont {Jozsa},\ and\
  \citenamefont {Schumacher}}]{Barnum1996}%
  \BibitemOpen
  \bibfield  {author} {\bibinfo {author} {\bibfnamefont {H.}~\bibnamefont
  {Barnum}}, \bibinfo {author} {\bibfnamefont {C.~M.}\ \bibnamefont {Caves}},
  \bibinfo {author} {\bibfnamefont {C.~A.}\ \bibnamefont {Fuchs}}, \bibinfo
  {author} {\bibfnamefont {R.}~\bibnamefont {Jozsa}},\ and\ \bibinfo {author}
  {\bibfnamefont {B.}~\bibnamefont {Schumacher}},\ }\bibfield  {title}
  {\bibinfo {title} {{Noncommuting Mixed States Cannot Be Broadcast}},\ }\href
  {https://doi.org/10.1103/PhysRevLett.76.2818} {\bibfield  {journal} {\bibinfo
   {journal} {Phys. Rev. Lett.}\ }\textbf {\bibinfo {volume} {76}},\ \bibinfo
  {pages} {2818} (\bibinfo {year} {1996})}\BibitemShut {NoStop}%
\bibitem [{\citenamefont {Arthurs}\ and\ \citenamefont
  {Kelly~Jr.}(1965)}]{Arthurs1965}%
  \BibitemOpen
  \bibfield  {author} {\bibinfo {author} {\bibfnamefont {E.}~\bibnamefont
  {Arthurs}}\ and\ \bibinfo {author} {\bibfnamefont {J.~L.}\ \bibnamefont
  {Kelly~Jr.}},\ }\bibfield  {title} {\bibinfo {title} {{On the Simultaneous
  Measurement of a Pair of Conjugate Observables}},\ }\href
  {https://doi.org/https://doi.org/10.1002/j.1538-7305.1965.tb01684.x}
  {\bibfield  {journal} {\bibinfo  {journal} {Bell Syst. Tech. J.}\ }\textbf
  {\bibinfo {volume} {44}},\ \bibinfo {pages} {725} (\bibinfo {year}
  {1965})}\BibitemShut {NoStop}%
\bibitem [{\citenamefont {Cerf}\ and\ \citenamefont
  {Iblisdir}(2001)}]{Cerf2001}%
  \BibitemOpen
  \bibfield  {author} {\bibinfo {author} {\bibfnamefont {N.~J.}\ \bibnamefont
  {Cerf}}\ and\ \bibinfo {author} {\bibfnamefont {S.}~\bibnamefont
  {Iblisdir}},\ }\bibfield  {title} {\bibinfo {title} {{Phase conjugation of
  continuous quantum variables}},\ }\href
  {https://doi.org/10.1103/PhysRevA.64.032307} {\bibfield  {journal} {\bibinfo
  {journal} {Phys. Rev. A}\ }\textbf {\bibinfo {volume} {64}},\ \bibinfo
  {pages} {032307} (\bibinfo {year} {2001})}\BibitemShut {NoStop}%
\bibitem [{\citenamefont {Giovannetti}\ \emph {et~al.}(2014)\citenamefont
  {Giovannetti}, \citenamefont {García-Patrón}, \citenamefont {Cerf},\ and\
  \citenamefont {Holevo}}]{Giovannetti2014}%
  \BibitemOpen
  \bibfield  {author} {\bibinfo {author} {\bibfnamefont {V.}~\bibnamefont
  {Giovannetti}}, \bibinfo {author} {\bibfnamefont {R.}~\bibnamefont
  {García-Patrón}}, \bibinfo {author} {\bibfnamefont {N.~J.}\ \bibnamefont
  {Cerf}},\ and\ \bibinfo {author} {\bibfnamefont {A.~S.}\ \bibnamefont
  {Holevo}},\ }\bibfield  {title} {\bibinfo {title} {{Ultimate classical
  communication rates of quantum optical channels}},\ }\href
  {https://doi.org/10.1038/nphoton.2014.216} {\bibfield  {journal} {\bibinfo
  {journal} {Nat. Photonics}\ }\textbf {\bibinfo {volume} {8}},\ \bibinfo
  {pages} {796} (\bibinfo {year} {2014})}\BibitemShut {NoStop}%
\bibitem [{\citenamefont {Giovannetti}\ \emph {et~al.}(2004)\citenamefont
  {Giovannetti}, \citenamefont {Guha}, \citenamefont {Lloyd}, \citenamefont
  {Maccone},\ and\ \citenamefont {Shapiro}}]{Giovannetti2004}%
  \BibitemOpen
  \bibfield  {author} {\bibinfo {author} {\bibfnamefont {V.}~\bibnamefont
  {Giovannetti}}, \bibinfo {author} {\bibfnamefont {S.}~\bibnamefont {Guha}},
  \bibinfo {author} {\bibfnamefont {S.}~\bibnamefont {Lloyd}}, \bibinfo
  {author} {\bibfnamefont {L.}~\bibnamefont {Maccone}},\ and\ \bibinfo {author}
  {\bibfnamefont {J.~H.}\ \bibnamefont {Shapiro}},\ }\bibfield  {title}
  {\bibinfo {title} {{Minimum output entropy of bosonic channels: A
  conjecture}},\ }\href {https://doi.org/10.1103/PhysRevA.70.032315} {\bibfield
   {journal} {\bibinfo  {journal} {Phys. Rev. A}\ }\textbf {\bibinfo {volume}
  {70}},\ \bibinfo {pages} {032315} (\bibinfo {year} {2004})}\BibitemShut
  {NoStop}%
\bibitem [{\citenamefont {Garc\'{\i}a-Patr\'on}\ \emph
  {et~al.}(2012)\citenamefont {Garc\'{\i}a-Patr\'on}, \citenamefont
  {Navarrete-Benlloch}, \citenamefont {Lloyd}, \citenamefont {Shapiro},\ and\
  \citenamefont {Cerf}}]{Garcia-Patron2012}%
  \BibitemOpen
  \bibfield  {author} {\bibinfo {author} {\bibfnamefont {R.}~\bibnamefont
  {Garc\'{\i}a-Patr\'on}}, \bibinfo {author} {\bibfnamefont {C.}~\bibnamefont
  {Navarrete-Benlloch}}, \bibinfo {author} {\bibfnamefont {S.}~\bibnamefont
  {Lloyd}}, \bibinfo {author} {\bibfnamefont {J.~H.}\ \bibnamefont {Shapiro}},\
  and\ \bibinfo {author} {\bibfnamefont {N.~J.}\ \bibnamefont {Cerf}},\
  }\bibfield  {title} {\bibinfo {title} {{Majorization Theory Approach to the
  Gaussian Channel Minimum Entropy Conjecture}},\ }\href
  {https://doi.org/10.1103/PhysRevLett.108.110505} {\bibfield  {journal}
  {\bibinfo  {journal} {Phys. Rev. Lett.}\ }\textbf {\bibinfo {volume} {108}},\
  \bibinfo {pages} {110505} (\bibinfo {year} {2012})}\BibitemShut {NoStop}%
\end{thebibliography}%


\begin{thebibliography}{10}%
\makeatletter
\providecommand \@ifxundefined [1]{%
 \@ifx{#1\undefined}
}%
\providecommand \@ifnum [1]{%
 \ifnum #1\expandafter \@firstoftwo
 \else \expandafter \@secondoftwo
 \fi
}%
\providecommand \@ifx [1]{%
 \ifx #1\expandafter \@firstoftwo
 \else \expandafter \@secondoftwo
 \fi
}%
\providecommand \natexlab [1]{#1}%
\providecommand \enquote  [1]{``#1''}%
\providecommand \bibnamefont  [1]{#1}%
\providecommand \bibfnamefont [1]{#1}%
\providecommand \citenamefont [1]{#1}%
\providecommand \href@noop [0]{\@secondoftwo}%
\providecommand \href [0]{\begingroup \@sanitize@url \@href}%
\providecommand \@href[1]{\@@startlink{#1}\@@href}%
\providecommand \@@href[1]{\endgroup#1\@@endlink}%
\providecommand \@sanitize@url [0]{\catcode `\\12\catcode `\$12\catcode
  `\&12\catcode `\#12\catcode `\^12\catcode `\_12\catcode `\%12\relax}%
\providecommand \@@startlink[1]{}%
\providecommand \@@endlink[0]{}%
\providecommand \url  [0]{\begingroup\@sanitize@url \@url }%
\providecommand \@url [1]{\endgroup\@href {#1}{\urlprefix }}%
\providecommand \urlprefix  [0]{URL }%
\providecommand \Eprint [0]{\href }%
\providecommand \doibase [0]{https://doi.org/}%
\providecommand \selectlanguage [0]{\@gobble}%
\providecommand \bibinfo  [0]{\@secondoftwo}%
\providecommand \bibfield  [0]{\@secondoftwo}%
\providecommand \translation [1]{[#1]}%
\providecommand \BibitemOpen [0]{}%
\providecommand \bibitemStop [0]{}%
\providecommand \bibitemNoStop [0]{.\EOS\space}%
\providecommand \EOS [0]{\spacefactor3000\relax}%
\providecommand \BibitemShut  [1]{\csname bibitem#1\endcsname}%
\let\auto@bib@innerbib\@empty
\bibitem [{\citenamefont {Hackl}\ and\ \citenamefont
  {Bianchi}(2021)}]{Hackl2021}%
  \BibitemOpen
  \bibfield  {author} {\bibinfo {author} {\bibfnamefont {L.}~\bibnamefont
  {Hackl}}\ and\ \bibinfo {author} {\bibfnamefont {E.}~\bibnamefont
  {Bianchi}},\ }\bibfield  {title} {\bibinfo {title} {{Bosonic and fermionic
  Gaussian states from Kähler structures}},\ }\href
  {https://doi.org/10.21468/SciPostPhysCore.4.3.025} {\bibfield  {journal}
  {\bibinfo  {journal} {SciPost Phys. Core}\ }\textbf {\bibinfo {volume} {4}},\
  \bibinfo {pages} {025} (\bibinfo {year} {2021})}\BibitemShut {NoStop}%
\bibitem [{\citenamefont {Cahill}\ and\ \citenamefont
  {Glauber}(1999)}]{Cahill1999}%
  \BibitemOpen
  \bibfield  {author} {\bibinfo {author} {\bibfnamefont {K.~E.}\ \bibnamefont
  {Cahill}}\ and\ \bibinfo {author} {\bibfnamefont {R.~J.}\ \bibnamefont
  {Glauber}},\ }\bibfield  {title} {\bibinfo {title} {{Density operators for
  fermions}},\ }\href {https://doi.org/10.1103/PhysRevA.59.1538} {\bibfield
  {journal} {\bibinfo  {journal} {Phys. Rev. A}\ }\textbf {\bibinfo {volume}
  {59}},\ \bibinfo {pages} {1538} (\bibinfo {year} {1999})}\BibitemShut
  {NoStop}%
\bibitem [{\citenamefont {Bröcker}\ and\ \citenamefont
  {Werner}(1995)}]{Broeckner1995}%
  \BibitemOpen
  \bibfield  {author} {\bibinfo {author} {\bibfnamefont {T.}~\bibnamefont
  {Bröcker}}\ and\ \bibinfo {author} {\bibfnamefont {R.~F.}\ \bibnamefont
  {Werner}},\ }\bibfield  {title} {\bibinfo {title} {{Mixed states with
  positive Wigner functions}},\ }\href {https://doi.org/10.1063/1.531326}
  {\bibfield  {journal} {\bibinfo  {journal} {J. Math. Phys.}\ }\textbf
  {\bibinfo {volume} {36}},\ \bibinfo {pages} {62} (\bibinfo {year}
  {1995})}\BibitemShut {NoStop}%
\bibitem [{\citenamefont {Mandilara}\ \emph {et~al.}(2009)\citenamefont
  {Mandilara}, \citenamefont {Karpov},\ and\ \citenamefont
  {Cerf}}]{Mandilara2009}%
  \BibitemOpen
  \bibfield  {author} {\bibinfo {author} {\bibfnamefont {A.}~\bibnamefont
  {Mandilara}}, \bibinfo {author} {\bibfnamefont {E.}~\bibnamefont {Karpov}},\
  and\ \bibinfo {author} {\bibfnamefont {N.~J.}\ \bibnamefont {Cerf}},\
  }\bibfield  {title} {\bibinfo {title} {{Extending Hudson's theorem to mixed
  quantum states}},\ }\href {https://doi.org/10.1103/PhysRevA.79.062302}
  {\bibfield  {journal} {\bibinfo  {journal} {Phys. Rev. A}\ }\textbf {\bibinfo
  {volume} {79}},\ \bibinfo {pages} {062302} (\bibinfo {year}
  {2009})}\BibitemShut {NoStop}%
\bibitem [{\citenamefont {Van~Herstraeten}\ and\ \citenamefont
  {Cerf}(2021)}]{VanHerstraeten2021a}%
  \BibitemOpen
  \bibfield  {author} {\bibinfo {author} {\bibfnamefont {Z.}~\bibnamefont
  {Van~Herstraeten}}\ and\ \bibinfo {author} {\bibfnamefont {N.~J.}\
  \bibnamefont {Cerf}},\ }\bibfield  {title} {\bibinfo {title} {{Quantum Wigner
  entropy}},\ }\href {https://doi.org/10.1103/PhysRevA.104.042211} {\bibfield
  {journal} {\bibinfo  {journal} {Phys. Rev. A}\ }\textbf {\bibinfo {volume}
  {104}},\ \bibinfo {pages} {042211} (\bibinfo {year} {2021})}\BibitemShut
  {NoStop}%
\bibitem [{\citenamefont {Lieb}\ and\ \citenamefont
  {Solovej}(2014)}]{Lieb2014b}%
  \BibitemOpen
  \bibfield  {author} {\bibinfo {author} {\bibfnamefont {E.~H.}\ \bibnamefont
  {Lieb}}\ and\ \bibinfo {author} {\bibfnamefont {J.~P.}\ \bibnamefont
  {Solovej}},\ }\bibfield  {title} {\bibinfo {title} {{Proof of an entropy
  conjecture for Bloch coherent spin states and its generalizations}},\ }\href
  {https://doi.org/10.1007/s11511-014-0113-6} {\bibfield  {journal} {\bibinfo
  {journal} {Acta Math.}\ }\textbf {\bibinfo {volume} {212}},\ \bibinfo {pages}
  {379} (\bibinfo {year} {2014})}\BibitemShut {NoStop}%
\bibitem [{\citenamefont {Wehrl}(1978)}]{Wehrl1978}%
  \BibitemOpen
  \bibfield  {author} {\bibinfo {author} {\bibfnamefont {A.}~\bibnamefont
  {Wehrl}},\ }\bibfield  {title} {\bibinfo {title} {{General properties of
  entropy}},\ }\href {https://doi.org/10.1103/RevModPhys.50.221} {\bibfield
  {journal} {\bibinfo  {journal} {Rev. Mod. Phys.}\ }\textbf {\bibinfo {volume}
  {50}},\ \bibinfo {pages} {221} (\bibinfo {year} {1978})}\BibitemShut
  {NoStop}%
\bibitem [{\citenamefont {Wehrl}(1979)}]{Wehrl1979}%
  \BibitemOpen
  \bibfield  {author} {\bibinfo {author} {\bibfnamefont {A.}~\bibnamefont
  {Wehrl}},\ }\bibfield  {title} {\bibinfo {title} {{On the relation between
  classical and quantum-mechanical entropy}},\ }\href
  {https://doi.org/10.1016/0034-4877(79)90070-3} {\bibfield  {journal}
  {\bibinfo  {journal} {Rep. Math. Phys.}\ }\textbf {\bibinfo {volume} {16}},\
  \bibinfo {pages} {353} (\bibinfo {year} {1979})}\BibitemShut {NoStop}%
\bibitem [{\citenamefont {Lieb}(1978)}]{Lieb1978}%
  \BibitemOpen
  \bibfield  {author} {\bibinfo {author} {\bibfnamefont {E.~H.}\ \bibnamefont
  {Lieb}},\ }\bibfield  {title} {\bibinfo {title} {{Proof of an entropy
  conjecture of Wehrl}},\ }\href {https://doi.org/10.1007/BF01940328}
  {\bibfield  {journal} {\bibinfo  {journal} {Commun. Math. Phys.}\ }\textbf
  {\bibinfo {volume} {62}},\ \bibinfo {pages} {35} (\bibinfo {year}
  {1978})}\BibitemShut {NoStop}%
\bibitem [{\citenamefont {Schupp}(2022)}]{Schupp2022}%
  \BibitemOpen
  \bibfield  {author} {\bibinfo {author} {\bibfnamefont {P.}~\bibnamefont
  {Schupp}},\ }\bibinfo {title} {{Wehrl entropy, coherent states and quantum
  channels}},\ in\ \href {https://doi.org/10.4171/90-2/42} {\emph {\bibinfo
  {booktitle} {{The Physics and Mathematics of Elliott Lieb}}}}\ (\bibinfo
  {publisher} {EMS Press},\ \bibinfo {year} {2022})\ pp.\ \bibinfo {pages}
  {329--344}\BibitemShut {NoStop}%
\end{thebibliography}%

\section{End Matter}

\textit{Appendix A: Cloning and the uncertainty principle}---The no-cloning theorem~\cite{Wootters1982,Dieks1982,Scarani2005} does not constrain a single fermion as the only two physical pure states $\ket{0}$ and $\ket{1}$ are orthogonal. Hence, a fermion can be perfectly cloned by employing a trivial $1 \to 2$ cloning strategy. Starting from a number state $\ket{m}$, we measure its occupation and prepare two independent fermions with the same mode population. This results in the state-independent average fidelity $F=\sum_{n,m=0}^1 p (m) F (n,m) = (1/2) \sum_{n=0}^1 (\delta_{n 0} + \delta_{n 1}) = 1$ for $p(m) = 1/2$. The unit fidelity for fermions is in stark contrast with other physical systems. The same trivial measure-and-prepare strategy yields $F=2/3$ for a single qubit (spin 1/2), which can be increased up to the optimal value $F=5/6$ by utilizing the universal quantum cloning machine proposed in~\cite{Buzek1996}, while the optimal Gaussian cloning strategy for a boson exhibits $F=2/3$~\cite{Cerf2000,Cerf2003}. More generally, a single fermion described by some physical density operator $\boldsymbol{\rho}$ can be broadcast, that is, copied to a second mode in the sense that the total state $\boldsymbol{\rho}_{A B}$ ends up with the marginals $\boldsymbol{\rho} = \Tr_A \{\boldsymbol{\rho}_{A B} \} = \Tr_B \{\boldsymbol{\rho}_{A B} \}$. As every physical state is diagonal in the Fock basis, it commutes with any other physical state, thereby bypassing the no-broadcasting theorem~\cite{Barnum1996}. 

A simple implementation of a fermionic broadcasting machine consists of the fermionic {\tt CNOT} gate. Given any physical input state $\boldsymbol{\rho}_{A}=(1-\braket{n})\boldsymbol{a} \boldsymbol{a}^{\dagger} + \braket{n} \boldsymbol{a}^{\dagger} \boldsymbol{a}$ on mode $A$ uncorrelated with the vacuum $\boldsymbol{\rho}_B = \boldsymbol{b} \boldsymbol{b}^{\dagger}$ on mode $B$, the {\tt CNOT} gate flips the occupation of mode $B$ if mode $A$ is populated and vice versa, which is described by the non-linear unitary $\boldsymbol{U} = (\boldsymbol{b} + \boldsymbol{b}^{\dagger}) \boldsymbol{a}^{\dagger} \boldsymbol{a} + \boldsymbol{a} \boldsymbol{a}^{\dagger}$. The corresponding canonical transformations $\boldsymbol{a} \to \boldsymbol{a} (\boldsymbol{b} +\boldsymbol{b}^\dagger ) $ and $\boldsymbol{b} \to \boldsymbol{a}^\dagger \boldsymbol{a} \boldsymbol{b}^\dagger + \boldsymbol{a} \boldsymbol{a}^\dagger \boldsymbol{b} $ conserve the anticommutation relations, leading to the two-mode output $\boldsymbol{\rho}_{A B}=(1-\braket{n})\boldsymbol{a} \boldsymbol{a}^{\dagger} \boldsymbol{b} \boldsymbol{b}^{\dagger} + \braket{n} \boldsymbol{a}^{\dagger} \boldsymbol{a} \boldsymbol{b}^{\dagger} \boldsymbol{b}$. The result is symmetric under the exchange of $A$ and $B$ and encodes the input state in both modes.

Interestingly, a simple analog of the Gaussian cloning machine for bosons~\cite{Cerf2000,Cerf2003} cannot exist for fermions. More precisely, there is no linear transformation that is covariant with respect to displacements, \textit{i.e.}, which conserves the mean. This can be shown as follows. Let $\boldsymbol{a} = (\boldsymbol{a}_1, \dots, \boldsymbol{a}_N)^T$ denote the input modes with mode $\boldsymbol{a}_1$ carrying the state to be cloned. The two clones shall appear in the output modes $\boldsymbol{b}_1$ and $\boldsymbol{b}_2$, with the output modes operators being described by $\boldsymbol{b} = \theta \boldsymbol{a} + \lambda \boldsymbol{a}^{\dagger}$ where $\theta, \lambda \in \mathbb{C}$ represent linear transformations. As the input state's mean shall be conserved, the two cloned output modes must read $\boldsymbol{b}_j = \boldsymbol{a}_1 + \sum_{k=2}^N (\theta_{j k} \boldsymbol{a}_k + \lambda_{j k} \boldsymbol{a}^{\dagger})$ for $j=1,2$. This, however, is impossible if all input and output modes are considered fermionic and independent, \textit{i.e.}, fulfill the anti-commutation relations $\{\boldsymbol{a}_j, \boldsymbol{a}_{j'}^{\dagger} \} = \{\boldsymbol{b}_j, \boldsymbol{b}_{j'}^{\dagger} \} = \delta_{j j'} \mathds{1}$, as this results in the contradiction $\{\boldsymbol{b}_1, \boldsymbol{b}_1^{\dagger} \} = \{\boldsymbol{a}_1, \boldsymbol{a}_1^\dagger \} + \sum_{k=2}^N (\abs{\theta_{j k}}^2 + \abs{\lambda_{j k}}^2) \{ \boldsymbol{a}_k, \boldsymbol{a}_k^\dagger \} \neq \mathds{1}$ whenever $\theta, \lambda \neq 0$. For bosons, the relative sign in the sum is a minus instead, and hence $\theta_{j k}$ and $\lambda_{j k}$ can be chosen such that the second term vanishes (yielding a linear cloning transformation). The fermionic {\tt CNOT} gate implementing perfect cloning is not affected by this no-go theorem as it is a nonlinear transformation.
    
For bosons, the impossibility of perfect cloning is expressed by the no-cloning uncertainty relations, which provide lower bounds on the minimum noise that a cloning machine adds to the (imperfect) clones~\cite{Cerf2000,Cerf2003}. We denote by $\boldsymbol{X} = \boldsymbol{x} + \tilde{\boldsymbol{x}}$ and $\boldsymbol{P} = \boldsymbol{p} + \tilde{\boldsymbol{p}}$ the quadratures being measured on two \textit{separate} clones stemming from the input mode $(\boldsymbol{x},\boldsymbol{p})$ and independent noise $(\tilde{\boldsymbol{x}}, \tilde{\boldsymbol{p}})$. As the joint measurement of $(\boldsymbol{X},\boldsymbol{P})$ is carried out over independent (bosonic) modes, we have $[\boldsymbol{X}, \boldsymbol{P}] = 0$, which implies $[\tilde{\boldsymbol{x}},\tilde{\boldsymbol{p}}] = - [\boldsymbol{x}, \boldsymbol{p}]=-i \mathds{1}$ since the noise is independent of the input, that is, $[ \boldsymbol{x}, \tilde{\boldsymbol{p}} ] = [ \tilde{\boldsymbol{x}},\boldsymbol{p} ] = 0$. As a consequence, $\sigma^2 (\tilde{\boldsymbol{x}}) \, \sigma^2 (\tilde{\boldsymbol{p}}) \ge 1/4$. Assuming equal noise in both quadratures, this shows that the minimum added noise to the input quadratures $(\boldsymbol{x},\boldsymbol{p})$ is one unit of shot noise, \textit{i.e.}, $\sigma^2 (\tilde{\boldsymbol{x}}) = \sigma^2 (\tilde{\boldsymbol{p}}) \ge 1/2$. This implies that the measured output quadratures $(\boldsymbol{X},\boldsymbol{P})$ are constrained by a relation of Arthurs-Kelly type, i.e.,  $\sigma^2 (\boldsymbol{X}) \, \sigma^2 (\boldsymbol{P}) \ge 1$~\cite{Arthurs1965} and hence suffer from twice the shot noise compared to the input-mode uncertainty. We remark that the Gaussian cloning machine saturates all these inequalities.

The uncertainty relations derived in the main text constrain the $xp$-correlations via $\sigma^2 (\boldsymbol{x}, \boldsymbol{p}) = \det \gamma (\boldsymbol{\rho}) = \det \gamma (W) \ge -1/4$, see also~\cite{SM}\textbf{(f)}. A joint measurement over \textit{separate} fermionic clones requires anti-commuting quadratures, \textit{i.e.}, $\{\boldsymbol{X}, \boldsymbol{P}\} = 0$, which implies $\{ \tilde{\boldsymbol{x}}, \tilde{\boldsymbol{p}} \} = 0$ since $\{ \boldsymbol{x}, \tilde{\boldsymbol{p}} \} = \{ \tilde{\boldsymbol{x}}, \boldsymbol{p} \} = 0$. At the same time, however, the noise shall be independent of the input $[ \boldsymbol{x}, \tilde{\boldsymbol{p}} ] = [ \tilde{\boldsymbol{x}},\boldsymbol{p} ] = 0$. The resulting equations $\boldsymbol{x} \tilde{\boldsymbol{p}} = \tilde{\boldsymbol{p}} \boldsymbol{x} = \tilde{\boldsymbol{x}}\boldsymbol{p} =\boldsymbol{p} \tilde{\boldsymbol{x}} = 0$ have non-trivial solutions for $( \tilde{\boldsymbol{x}}, \tilde{\boldsymbol{p}})$ only if their ranges fully lie in the kernels of $(\boldsymbol{x}, \boldsymbol{p})$. The latter are empty sets over the fermionic Hilbert space $\mathcal{H}_2 = \{ \ket{0}, \ket{1} \}$, implying $\tilde{\boldsymbol{x}} = \tilde{\boldsymbol{p}} = 0$. Thus, the output quadratures are constrained by the uncertainty principle in precisely the same way as the input quadratures, that is, $\sigma^2 (\boldsymbol{X}, \boldsymbol{P}) \ge -1/4$. In particular, they do \textit{not} suffer from additional noise -- in accordance with fermionic clones being perfect. Closely related is the possibility of phase conjugation: The conservation of anti-commutation relations in a fermionic transformation allows phase conjugation, $\boldsymbol{a} \to \boldsymbol{a}^\dagger$. This transformation is notoriously forbidden for bosons~\cite{Cerf2001}.

The situation changes drastically for two or more fermions as pure-state superpositions over various modes cannot be cloned perfectly. Nonetheless, it is always possible to separately clone all their single-mode marginals.

\smallskip
\textit{Appendix B: Fermionic channels in phase space}---In the continuous-variable quantum information formalism, information and majorization theories play important roles in characterizing Gaussian quantum channels.~\cite{Weedbrook2012,Serafini2017}. Here, we sketch the situation for fermions. Without loss of generality, we consider the single-mode thermal loss channel, which represents a Gaussian diffusion (or heat flow) process: Some (thermal) input state $\boldsymbol{\rho}_{\text{in}}$ with mode number $\braket{n_{\text{in}}}$ is mixed with an independent environmental mode $\boldsymbol{\rho}_{\text{env}}$ carrying $\braket{n_{\text{env}}}$ particles on a beamsplitter of transmittivity $\tau \in [0,1]$, \textit{i.e.}, $\boldsymbol{\rho}_{\text{in}} \to \boldsymbol{\rho}_{\text{out}} = \bTr \{ \boldsymbol{U}_{\text{BS}} (\boldsymbol{\rho}_{\text{in}} \otimes \boldsymbol{\rho}_{\text{env}}) \boldsymbol{U}_{\text{BS}}^{\dagger} \}$. Here, 
\begin{equation*}
    U_{\text{BS}} = \begin{pmatrix}
        \sqrt{\tau} & \sqrt{1-\tau} \\
        \sqrt{1-\tau} & - \sqrt{\tau}
    \end{pmatrix}
\end{equation*}
is the unitary matrix that describes the beamsplitter transformation on the two modes $(\boldsymbol{a}_{\text{in}}, \boldsymbol{a}_{\text{env}})$ in phase space. It is straightforward to check that the physical part of the output state is of the standard thermal form $\boldsymbol{\rho}_{\text{out}} = (1 - \braket{n_{\text{out}}}) \boldsymbol{a} \boldsymbol{a}^{\dagger} + \braket{n_{\text{out}}} \boldsymbol{a}^{\dagger} \boldsymbol{a}$, where $\braket{n_{\text{out}}} = \tau \braket{n_{\text{in}}} + (1-\tau)\braket{n_{\text{env}}}$, which agrees with the bosonic result (this is not at all evident given that the Pauli exclusion principle forbids more than one fermion per mode). Indeed, a careful calculation shows that $\braket{n_{\text{out}}} =  \tau \braket{n_{\text{in}}} (1-\braket{n_{\text{env}}})+ (1-\tau)\braket{n_{\text{env}}}(1-\braket{n_{\text{in}}}) + \braket{n_{\text{in}}} \braket{n_{\text{env}}}$. Since the set of possible output states comprises all physical states for any given input, any single-mode fermionic channel can be written as a thermal loss channel.
    
In spite of the Grassmannian nature of the displacement variable $\alpha$ (or, equivalently, of the $x$ and $p$ quadratures), it is formally possible to transpose the notions of Gaussian noise and modulation. Given any input phase space distribution $z_{\text{in}} (\alpha)$, the thermal loss channel yields the output distribution $z_{\text{out}} (\gamma) = \int \mathcal{D} \alpha \,  z_{\text{in}} (\alpha) \, z_{\tau, \braket{n_{\text{env}}}} (\gamma - \alpha)$, defined as a Grassmannian convolution. We determine the channel's kernel $z_{\tau, \braket{n_{\text{env}}}} (\alpha)$ using the Glauber $P$-representation. At the input, the state reads $\boldsymbol{\rho}_{\text{in}} = \int \mathcal{D} \alpha  \, P_{\text{in}} (\alpha) \boldsymbol{\ket{\alpha}{\bra{-\alpha}}}$ with $P_{\text{in}} (\alpha) =-\braket{n_{\text{in}}}+\alpha\alpha^*$, which is a Gaussian-modulated coherent state. Similarly, the environmental state reads $\boldsymbol{\rho}_{\text{env}} = \int \mathcal{D} \beta  \, P_{\text{env}} (\beta) \boldsymbol{\ket{\beta}{\bra{-\beta}}}$ with $P_{\text{env}} (\beta) =-\braket{n_{\text{env}}}+\beta\beta^*$. At the output, the state reads $\boldsymbol{\rho}_{\text{out}} = \int \mathcal{D} \alpha \,\mathcal{D} \beta \, P_{\text{in}} (\alpha) P_{\text{env}} (\beta) \boldsymbol{\ket{\gamma}{\bra{-\gamma}}}$, where $\gamma = \sqrt{\tau} \alpha + \sqrt{1-\tau} \beta$ according to the beam splitter transformation for (fermionic) coherent states. Substituting $\beta \to \gamma$ and recalling that we must include the \textit{inverse} of the Jacobian $J=1/(1-\tau)$ for Berezin integrals, we obtain $\boldsymbol{\rho}_{\text{out}} = \int \mathcal{D} \alpha \,\mathcal{D} \gamma \, J^{-1} \, P_{\text{in}} (\alpha) P_{\text{env}} [\gamma/\sqrt{1-\tau} - \alpha \sqrt{\tau/(1-\tau)}] \boldsymbol{\ket{\gamma}{\bra{-\gamma}}}$. Thus, comparing with the Glauber $P$-representation of the output state  $\boldsymbol{\rho}_{\text{out}} = \int \mathcal{D} \gamma \, P_{\text{out}} (\gamma) \boldsymbol{\ket{\gamma}{\bra{-\gamma}}}$ implies that the kernel is $P_{\tau, \braket{n_\text{env}}} = (1-\tau) P_{\text{env}} [\gamma/\sqrt{1-\tau} - \alpha \sqrt{\tau/(1-\tau)}]$, in accordance with the bosonic case: Gaussian noise is added, while coherent components are (unphysically) shifted.

It is instructive to notice that the relation between the input and output occupation numbers that was derived in the Fock basis can also be understood in phase space. Starting from $\boldsymbol{\rho}_{\text{out}} = \int \mathcal{D} \alpha \,\mathcal{D} \beta \, (-\braket{n_{\text{in}}}+\alpha\alpha^*) (-\braket{n_{\text{env}}}+\beta\beta^*) \boldsymbol{\ket{\gamma}{\bra{-\gamma}}}$, with $\gamma = \sqrt{\tau} \alpha + \sqrt{1-\tau} \beta$ and transforming $\beta \to \gamma$, we may integrate over $\alpha$, yielding $\boldsymbol{\rho}_{\text{out}} = \int \mathcal{D} \gamma \, (-\tau \braket{n_{\text{in}}} - (1-\tau)\braket{n_{\text{env}}}+\gamma\gamma^*) \boldsymbol{\ket{\gamma}{\bra{-\gamma}}}$, confirming the relation $\braket{n_{\text{out}}} = \tau \braket{n_{\text{in}}} + (1-\tau)\braket{n_{\text{env}}}$. Thus, the convolution between the Gaussian-modulated input and Gaussian-distributed noise results in a Gaussian-distributed output, in full analogy with the bosonic case (the coherent states are unphysical, but their Gaussian mixture coincides with a mixture of $\ket{0}$ and $\ket{1}$).

Our fermionic majorization theory in phase space states that the disorder of a single fermionic mode grows monotonically from the vacuum (least disorder) to the excited state (most disorder). Indeed, when inserting the vacuum, the thermal loss channel adds thermal noise (corresponding to an inflow of heat) and thus increases the system's disorder as $\braket{n_{\text{out}}} = (1-\tau)\braket{n_{\text{env}}} \ge \braket{n_{\text{in}}} = 0$. More generally, some input distribution majorizes its output distribution, \textit{i.e.}, $z_{\text{out}} \prec z_{\text{in}}$, if the environment is hotter than the input, \textit{i.e.}, $\braket{n_{\text{env}}} > \braket{n_{\text{in}}}$ (vice versa if the latter condition is reversed). This is also reflected by the behavior of the Wehrl entropy (which extends the classical Boltzmann entropy to the quantum level and, as such, captures thermal properties~\cite{Wehrl1978}) when considered as a function of $\tau$ and $\braket{n_{\text{env}}}$. At the same time, uncertainty increases as well. Note, however, that the convenient measure for uncertainty -- the Wigner covariance matrix' determinant -- is monotonic in $\tau$ and $\braket{n_{\text{env}}}$ only if considered separately over $\braket{n_{\text{env}}} \ge [2(1-\tau )]^{-1}$ or the complementary region. The uncertainty measure attains a global maximum when the latter inequality becomes tight, which marks the point at which the output mode's temperature is either positive or negative.

A central question in the information-theoretic analysis of communication concerns the (classical) capacity of a channel, that is, its maximum communication rate and how it can be reached. In the bosonic case, it is well-known that this is achieved by coherent-state inputs~\cite{Giovannetti2014} (see also~\cite{Giovannetti2004,Garcia-Patron2012}). This originates from the key fact that the output von Neumann entropy attains its minimum when the input state is coherent. When considering fermions, we may equally work with the Wigner entropy, as it is monotonically increasing (decreasing) for $\braket{n} < (>) 1/2$, just as the von Neumann entropy. The output Wigner entropy of a thermal loss channel is $S (W_{\text{out}}) = -1 - \ln \abs{1/2 - \braket{n_{\text{out}}}}$. For given $\tau$ and $\braket{n_{\text{env}}}$, the minimum is always obtained for a pure input state (which also follows from the Wigner entropy being convex over each temperature branch). More precisely, the vacuum (excited state) input is optimal when $\braket{n_{\text{env}}} < (>) 1/2$.

\end{document}


\title{Supplementary material}

\maketitle

Here, we provide details on the fermionic formalism and additional remarks for the interested reader. \\

\textbf{(a) (Un)physical states}

\begin{itemize}
    \item When allowing $\lambda \neq 0$, the state in Eq.~(\textcolor{blue}{1}) is still Gaussian, \textit{i.e.}, it can be written as $\boldsymbol{\rho} = \exp (-2 \braket{n} \boldsymbol{a}^{\dagger} \boldsymbol{a} + \lambda \boldsymbol{a} + \lambda^* \boldsymbol{a}^{\dagger} + \braket{n} - \abs{\lambda}^2)$, but it is neither thermal, nor physical (as we shall see, this actually corresponds to a Gaussian state that is not centered at the origin).
\end{itemize}

\textbf{(b) Berezin integrals} 

\begin{itemize}
    \item Berezin integrals are integrals over Grassmann-valued variables, which are such that $\int  \mathrm{d} \alpha \, 1 = 0$ and $\int  \mathrm{d} \alpha \, \alpha = 1$, showing that integration is equivalent to differentiation for Grassmann variables. Integrals of all other powers vanish since $\alpha^n = 0$ for $n\ge 2$. 
\end{itemize}

\textbf{(c) Fermionic coherent states} 

\begin{itemize}
    \item Fermionic coherent states $\ket{\alpha}$ are characterized by Grassmann-valued displacements and thus are unphysical unless $\alpha=0$, see also~\cite{Hackl2021}. Indeed, using Eq.~(\textcolor{blue}{3}) together with $\boldsymbol{D}(\alpha) = \mathds{1} + \boldsymbol{a}^{\dagger} \alpha - \alpha^* \boldsymbol{a} + \left( \tfrac{1}{2} \,\mathds{1}- \boldsymbol{a}^{\dagger} \boldsymbol{a} \right) \alpha \alpha^*$, we see that coherent states are superpositions $\ket{\alpha}=(1+\frac{1}{2}\alpha\alpha^*)\ket{0}-\alpha\ket{1}$. In analogy to the bosonic case, fermionic coherent states are eigenstates of the annihilation operator $\boldsymbol{a} \ket{\alpha} = \alpha \ket{\alpha}$.

    \item The decomposition of the identity in terms of fermionic coherent states, namely $\mathds{1} = \int \mathcal{D} \alpha \boldsymbol{\ket{\alpha} \bra{\alpha}}$, can be checked by using $\ket{\alpha}= \beta \ket{0}-\alpha\ket{1}$ and $\bra{\alpha}= \beta^* \bra{0}-\alpha^*\bra{1}$, with $\beta =1+\frac{1}{2}\alpha\alpha^*$, together with the fact that $\int \mathcal{D} \alpha \, \alpha\alpha^* = \int \mathcal{D} \alpha \, \beta\beta^* = 1$ as well as $\int \mathcal{D} \alpha \, \alpha\beta^* = \int \mathcal{D} \alpha \, \beta\alpha^* = 0$. 

    \item The expression for the trace in the fermionic coherent basis, namely $\bTr \{ \boldsymbol{O} \} = \int \mathcal{D} \alpha \braket{\alpha | \boldsymbol{O} | - \alpha}$, can also be checked by using $\ket{-\alpha}= \beta \ket{0}+\alpha\ket{1}$ and $\bra{\alpha}= \beta^* \bra{0}-\alpha^*\bra{1}$, with $\beta =1+\frac{1}{2}\alpha\alpha^*$, together with the fact that $\int \mathcal{D} \alpha \, \beta^*\beta = \int \mathcal{D} \alpha \, (-\alpha^*)\alpha = 1$ as well as $\int \mathcal{D} \alpha \, \beta^*\alpha = \int \mathcal{D} \alpha \, (-\alpha^*)\beta = 0$. We may easily verify that the minus sign in $\ket{-\alpha}$ is needed by applying the trace formula to the density operator $\boldsymbol{\rho}$. Indeed, we get $\bTr \{ \boldsymbol{\rho} \} = 1$ whereas we would get $\bTr \{ \boldsymbol{\rho} \} = 1-2\braket{n}$ by (wrongly) replacing $\ket{-\alpha}$ with $\ket{\alpha}$, so that the excited state would (wrongly) be normalized to $-1$. 
    This issue is intrinsically linked to the fact that $\braket{n|\alpha}$ and $\braket{\alpha|n}$ anticommute for $n=1$ (whereas they commute for $n=0$), so we have to insert this minus sign when exchanging the two matrix elements (this fixes the problem both for $n=0$ and $n=1$).    
\end{itemize}

\textbf{(d) Fermionic phase-space distributions} 

\begin{itemize}
    \item The normalization of the Glauber $P$-distribution $\int \mathcal{D} \alpha \, P (\alpha) = \bTr \{\boldsymbol{\rho}\} =  1$ is ensured by the normalized projector $\bTr \{ \boldsymbol{\ket{\alpha} \bra{-\alpha}} \} = \braket{\alpha | \alpha} = 1$, whereas $\bTr \{ \boldsymbol{\ket{\alpha} \bra{\alpha}} \} = \braket{-\alpha | \alpha} = 1 - 2 \alpha^*\alpha$. Analogously, $ \int \mathcal{D} \alpha \, Q (\alpha)= \bTr \{\boldsymbol{\rho}\} = 1$ is a consequence of the minus sign appearing in the coherent-state trace formula.

    \item The characteristic function of $\boldsymbol{\rho}$ is calculated using $\bTr \{ \boldsymbol{\ket{0}\bra{0}} \boldsymbol{D}(\alpha) \}=1+\frac{1}{2}\alpha\alpha^*$ as well as $\bTr \{ \boldsymbol{\ket{1} \bra{1}} \boldsymbol{D}(\alpha) \}=1-\frac{1}{2}\alpha\alpha^*$, resulting in $\chi(\alpha)=1+(\frac{1}{2}-\braket{n})\alpha\alpha^*$. To get $W(\alpha)$, we perform the Fourier transform of $\chi(\alpha)$ by exploiting the identity $e^{\alpha \beta^* - \beta \alpha^*}=1+\alpha \beta^* - \beta \alpha^*+\alpha\alpha^*\beta\beta^*$. The expressions of $P (\alpha)$ and $Q(\alpha)$ are also straightforward to derive by using the expansions of $\boldsymbol{\rho}$ and $\ket{\alpha}$ in the Fock basis.

    \item We observe all phase-space distributions to be Grassmann-even. More generally, we argue that any physical quantity has to be of definite Grassmann-parity, that is, either fully commute or anti-commute with any Grassmann variable, which we refer to as Grassmann-even or Grassmann-odd, respectively (see also~\cite{Cahill1999}). 

    \item In contrast to the bosonic case, where a complete characterization of the set of Wigner-positive states is an outstanding problem~\cite{Broeckner1995,Mandilara2009,VanHerstraeten2021a}, the sign of the single-mode fermionic Wigner $W$-distribution is entirely determined by the particle number $\braket{n}$ and thus by the sign of the temperature. Recall that negative temperatures can occur for a Hamiltonian bounded from above when the occupation of the excited states is more likely. In our case, this amounts to $\braket{n}>1/2$. Also, the fermionic Glauber $P$-distribution is always a real function, while its bosonic counterpart can become distribution-valued involving the Dirac $\delta$-distribution and its derivatives.
\end{itemize}

\textbf{(e) Majorization relations} 

\begin{itemize}
    \item Without the condition $f(0)=0$ both sides of the majorization relation would diverge~\cite{Lieb2014b}.

    \item Although the proofs of the majorizations relations used the first derivative of $f$, we do not even have to assume that $f$ is analytic since all higher derivatives of $f$ are multiplied by zero. Further, the existence of its first derivative $f'$ is guaranteed almost everywhere following Rademacher's theorem for Lipschitz-continuous functions, with the exceptions occurring only at the boundary points.

    \item The equivalence between second-moment and majorization relations can be proven as follows. Any physical distribution $z_1 (\alpha_1)$ is related to another distribution $z_2 (\alpha_2)$ by $\alpha_2 \to \alpha_1 = \sqrt{M} \, \alpha_2$ such that $z_2 (\alpha_2) \to z_1 (\alpha_1) = M \, z_2 (\alpha_2)$, where $M = z_{2,\text{B}}/z_{1,\text{B}}$ is a real number. Then, the majorization relation Eq.~(\textcolor{blue}{8}) can be rewritten as $\int \mathcal{D} \alpha_1 \, f(z_1) = \int \mathcal{D} \alpha_2 \, f(M \, z_2)/M \ge \int \mathcal{D} \alpha_2 \, f(z_2)$. (Note here that a change of coordinates for Grassmann variables is accompanied by the \textit{inverse} Jacobian.) Since $f$ is concave and fulfills $f(0)=0$, it is subadditive. Thus, the latter relation is fulfilled if $0 \le M \le 1$ which implies $\det \gamma (z_1) \ge (\le) \det \gamma (z_2)$ when $\abs{z_{1,\text{B}}} \ge (\le) \abs{z_{2,\text{B}}}$. The converse statement follows from transforming the left-hand side of Eq.~(\textcolor{blue}{8}) instead.
\end{itemize}

\textbf{(f) Moments in state space and phase space}

\begin{itemize}
    \item To relate quantities in state space and phase space, we introduce the fermionic variant of the overlap formula for two physical density operators, which reads $\bTr \{ \boldsymbol{\rho}_1 \boldsymbol{\rho}_2 \} = \int \mathcal{D} \alpha \, W_1 (\alpha) K_2 (\alpha)$, where $W_1(\alpha)=1/2 -  \braket{n_1}+ \alpha \alpha^*$ and 
    $K_2 (\alpha) = 1/2 + (1 - 2 \braket{n_2}) \alpha \alpha^*$, with $\braket{n_1}$ ($\braket{n_2}$) denoting the mean particle number in $\boldsymbol{\rho}_1$ ($\boldsymbol{\rho}_2$).
    
    \item The mean fields of any physical state vanish $\bTr \{ \boldsymbol{\rho} \boldsymbol{a} \} = \int \mathcal{D} \alpha \, W(\alpha) \, \alpha^*=0$ and $\bTr \{ \boldsymbol{\rho} \boldsymbol{a}^{\dagger} \} = \int \mathcal{D} \alpha \, W(\alpha) \, \alpha = 0$. Hence, physical states are Gaussian states that are centered at the origin, see also Eq.~(\textcolor{blue}{2}), and thus all second-order moments $\gamma_{j j'} (z)$ are centered moments. 

    \item By setting $\braket{n_2}$ to either $0$ or $1$, the overlap formula shows that the state's covariance matrix defined as $\gamma_{j j'} (\boldsymbol{\rho}) \equiv \bTr \{ \boldsymbol{\rho} [\boldsymbol{\xi}_j, \boldsymbol{\xi}_{j'}] \}/(2i)$, where $\boldsymbol{\xi}_1 = \boldsymbol{a}$ and $\boldsymbol{\xi}_2 = \boldsymbol{a}^{\dagger}$, agrees with the one of the Wigner $W$-distribution, \textit{i.e.}, $\gamma (\boldsymbol{\rho}) = \gamma (W)$. More precisely, we find for the off-diagonal elements $\bTr \{ \boldsymbol{\rho} [\boldsymbol{a}, \boldsymbol{a}^{\dagger}] \} = \int \mathcal{D} \alpha \, W(\alpha) [\alpha ,\alpha^*]$, although similar relations do \textit{not} hold for the individual terms, consider, \textit{e.g.}, $\bTr \{ \boldsymbol{\rho} \boldsymbol{a} \boldsymbol{a}^{\dagger} \} = \int \mathcal{D} \alpha \, W(\alpha) (1/2 + \alpha \alpha^*) = 1 - \braket{n}$.

    \item The covariance matrix is also often defined in terms of the field quadratures. In contrast to bosons, fermionic quadratures $\boldsymbol{x} = (\boldsymbol{a} + \boldsymbol{a}^{\dagger})/\sqrt{2}$ and $\boldsymbol{p} = (\boldsymbol{a} - \boldsymbol{a}^{\dagger})/(\sqrt{2} i)$, which are commonly referred to as Majorana operators, anti-commute $\{ \boldsymbol{x}, \boldsymbol{p} \} = 0$ and square to a constant $\boldsymbol{x}^2 = \boldsymbol{p}^2 = \mathds{1}/2$. Noting that $[\boldsymbol{x},\boldsymbol{p}] = i[\boldsymbol{a},\boldsymbol{a}^{\dagger}]$ shows that the covariance matrix takes the form $\gamma_{j j'} (\boldsymbol{\rho}) = \bTr \{ \boldsymbol{\rho} [\boldsymbol{\xi}_j, \boldsymbol{\xi}_{j'}] \}/2$, where now $\boldsymbol{\xi}_1 = \boldsymbol{x}$ and $\boldsymbol{\xi}_2 = \boldsymbol{p}$. We note that the diagonal elements are still zero since every operator commutes with itself. Hence, the variances defined via $\sigma^2 (\boldsymbol{x}) \equiv \braket{\boldsymbol{x}^2} - \braket{\boldsymbol{x}}^2$ (which are constant $\sigma^2 (\boldsymbol{x}) = \sigma^2 (\boldsymbol{p}) = 1/2$) do not appear in the covariance matrix. Instead, $ \gamma (\boldsymbol{\rho}) = i \sigma (\boldsymbol{x}, \boldsymbol{p}) \sigma_y$, where $\sigma (\boldsymbol{x}, \boldsymbol{p}) \equiv \braket{[\boldsymbol{x}, \boldsymbol{p}]}/2$ denotes the anti-symmetrized covariance and $\sigma_y$ is the second Pauli matrix. Hence, we have $\det \gamma (\boldsymbol{\rho}) =\sigma^2 (\boldsymbol{x}, \boldsymbol{p})= \braket{[\boldsymbol{x}, \boldsymbol{p}]}^2/4$. 
\end{itemize}

\textbf{(g) Entropies and entropic uncertainty relations}

\begin{itemize}
    \item Continuous-variable entropies of classical probability distributions are commonly denoted by $h(\cdot)$. Still, we choose the convention $S(\cdot)$ that is preferred in the mathematical literature, see, \textit{e.g.},~\cite{Wehrl1978,Wehrl1979,Lieb1978,Lieb2014b,Schupp2022}.

    \item Defining entropies with an additional minus sign to render them concave would make them lose their meaning as uncertainty measures, since, \textit{e.g.}, $S(W)$ would then tend to $- \infty$ when approaching the maximally mixed state $\braket{n} \to 1/2$.

    \item For a comparison of the entropic uncertainty relations to the bosonic case, we note that the corresponding inequalities for a bosonic mode read $S_r(W)\ge (r-1)^{-1}\ln r+\ln \pi$ and $S_r(Q)\ge (r-1)^{-1}\ln r$. Since the fermionic Wigner $W$-distribution comes without the normalization factor of $2\pi$ that prevails in the bosonic case, the corrected bosonic lower bound on $S_r(W)$ would read  $(r-1)^{-1}\ln r - \ln 2$, which is precisely the fermionic lower bound with an overall minus sign.
\end{itemize}


\bibliography{references.bib}